%% file: main.tex
\documentclass[fleqn,usenatbib,useAMS]{mnras}

\usepackage{xcolor}

\usepackage{graphicx, float}	% Including figure files
\usepackage{amsmath}	% Advanced maths commands
\usepackage{amssymb}	% Extra maths symbols
\usepackage{multicol}        % Multi-column entries in tables
\usepackage{bm}		% Bold maths symbols, including upright Greek
\usepackage{pdflscape}	% Landscape pages
\usepackage{lipsum}
\usepackage{tipa}

\usepackage[T1]{fontenc}
\usepackage{ae,aecompl}

\usepackage{newtxtext,newtxmath}
\title[SALVAGE II: the central molecular gas of AGN hosts]{SALVAGE II: no systematic molecular gas depletion in the central 2 kpc of AGN hosts}

\author[S. Wilkinson et al.]{Scott Wilkinson$^{1}$\thanks{Contact e-mail: \href{mailto:swilkinson@uvic.ca}{swilkinson@uvic.ca}}, Chiara Circosta$^{2, 3}, $Toby Brown$^{4,1}$, Sara L. Ellison$^{1}$, Shoshannah Byrne-Mamahit$^{1}$,
\newauthor Almudena Alonso-Herrero$^{5}$, Santiago Garcia-Burillo$^{6}$, Federico Esposito$^{6}$
\\
$^{1}$Department of Physics and Astronomy, University of Victoria, 3800 Finnerty Road, Victoria, BC V8P 5C2, Canada
\\
$^{2}$ESA, European Space Astronomy Centre (ESAC), Camino Bajo del Castillo s/n, 28692 Villanueva de la Cañada, Madrid, Spain
\\
$^{3}$Institut de Radioastronomie Millimétrique (IRAM), 300 rue de la Piscine, 38400 Saint-Martin-d’Hères, France
\\
$^{4}$Herzberg Astronomy and Astrophysics Research Centre, National Research Council of Canada, 5071 West Saanich Rd, Victoria, BC V9E 2E7, Canada
\\
$^{5}$Centro de Astrobiología (CAB), CSIC-INTA, Camino Bajo del Castillo s/n, 28692 Villanueva de la Cañada, Madrid, Spain
\\
$^{6}$Observatorio Astronómico Nacional (OAN-IGN)-Observatorio de Madrid, Alfonso XII, 3, 28014-Madrid, Spain
}

\date{September 21st, 2026}

\pubyear{2026}

\begin{document}
\label{firstpage}
\pagerange{\pageref{firstpage}--\pageref{lastpage}}
\maketitle

\begin{abstract}

In cosmological simulations, active galactic nuclei (AGN) feedback is found to be necessary for the shutdown of galaxy-wide star-formation by heating and/or removing molecular gas. However, at low-redshift, there is little evidence of systematic large-scale gas depletion. Some recent observations have found molecular gas depletion in AGN hosts, but limited to the central regions of galaxies. In this work, we use the SDSS-ALMA Legacy-Value Archival Gas Exploration (SALVAGE) dataset, which provides molecular gas  and stellar mass measurements from the inner ($r < $ 0.5-2 kpc) and outer region of the galaxy. Using AGN diagnostics from optical emission lines, mid-IR colour, and X-ray luminosity, we identify 70 AGN at $z < 0.07$ and inspect their central molecular gas reservoirs with a suite of metrics compared to non-AGN controls. In agreement with global studies of low-$z$ AGN hosts, we find that, compared to matched non-AGN, the central regions of AGN have equal molecular gas fractions and enhanced molecular gas surface densities. For the first time, we explore the central gas reservoirs of AGN identified with different selection criteria and find two optical AGN classes that show a tentative $\thicksim1.5\sigma$ and $\thicksim1.8\sigma$ signal of central molecular gas depletion. However, the AGN with central gas depletion do not occupy any specific region of the parameter space, such as AGN luminosity, that can provide a physical explanation for the depletion. If AGN are responsible for large-scale gas removal, the lack of systematic gas depletion in SALVAGE AGN sets an upper limit on the physical scale of the gas removal while the AGN remains observable.

\end{abstract}

%We conclude that either AGN do not systematically deplete central molecular gas reservoirs or 2 kpc resolution is too large to 

\begin{keywords}
galaxies: active, galaxies: nuclei, galaxies: ISM, galaxies: evolution, submillimetre: galaxies
\end{keywords}

\input{S1_Introduction}

\input{S2_Data}
\input{S3_Sample}
\input{S4_Results}

\input{S5_Discussion}

\input{S6_Summary}

\section*{Acknowledgements}

We respectfully acknowledge the L\textschwa\textvbaraccent {k}$^{\rm w}$\textschwa\ng{}\textschwa n Peoples on whose traditional territory the University of Victoria stands and the Songhees, Esquimalt and $\underline{\text{W}}\acute{\text{S}}$ANE$\acute{\text{C}}$ peoples whose relationships with the land continue to this day. As we explore the shared sky, we acknowledge our responsibilities to honour those who were here before us, and their continuing relationships to these lands. We strive for respectful relationships and partnerships with all the peoples of these lands as we move forward together towards reconciliation and decolonization. 

SW gratefully acknowledges the support from the Natural Sciences and Engineering Council of Canada (NSERC) as part of their graduate fellowship program. SLE gratefully acknowledges the receipt of NSERC Discovery Grants. Cette recherche a été financée par le Conseil de recherches en sciences naturelles et en génie du Canada (CRSNG).

SW and CC acknowledge support from the ESA Science Faculty Visitor Funding, reference ESA-SCI-E-LE-094. SGB acknowledges support from the Spanish grant PID2022-138560NB-I00, funded by MCIN/AEI/10.13039/501100011033/FEDER, EU.

The authors acknowledge the use of the Canadian Advanced Network for Astronomy Research (CANFAR) Science Platform. Our work used the facilities of the Canadian Astronomy Data Center, operated by the National Research Council of Canada with the support of the Canadian Space Agency, and CANFAR, a consortium that serves the data-intensive storage, access, and processing needs of university groups and centers engaged in astronomy research.

This research used the Canadian Advanced Network For Astronomy Research (CANFAR) operated in partnership by the Canadian Astronomy Data Centre and The Digital Research Alliance of Canada with support from the National Research Council of Canada the Canadian Space Agency, CANARIE and the Canadian Foundation for Innovation.

Funding for the SDSS and SDSS-II has been provided by the Alfred P. Sloan Foundation, the Participating Institutions, the National Science Foundation, the U.S. Department of Energy, the National Aeronautics and Space Administration, the Japanese Monbukagakusho, the Max Planck Society, and the Higher Education Funding Council for England. The SDSS Web Site is http://www.sdss.org/.

The SDSS is managed by the Astrophysical Research Consortium for the Participating Institutions. The Participating Institutions are the American Museum of Natural History, Astrophysical Institute Potsdam, University of Basel, University of Cambridge, Case Western Reserve University, University of Chicago, Drexel University, Fermilab, the Institute for Advanced Study, the Japan Participation Group, Johns Hopkins University, the Joint Institute for Nuclear Astrophysics, the Kavli Institute for Particle Astrophysics and Cosmology, the Korean Scientist Group, the Chinese Academy of Sciences (LAMOST), Los Alamos National Laboratory, the Max-Planck-Institute for Astronomy (MPIA), the Max-Planck-Institute for Astrophysics (MPA), New Mexico State University, Ohio State University, University of Pittsburgh, University of Portsmouth, Princeton University, the United States Naval Observatory, and the University of Washington.

This paper makes use of the following ALMA data: \\
ADS/JAO.ALMA\#2011.0.00374.S, ADS/JAO.ALMA\#2012.1.00539.S, ADS/JAO.ALMA\#2013.1.00058.S, ADS/JAO.ALMA\#2013.1.00096.S, ADS/JAO.ALMA\#2013.1.00115.S, ADS/JAO.ALMA\#2013.1.00530.S, ADS/JAO.ALMA\#2013.1.01383.S, ADS/JAO.ALMA\#2015.1.00320.S, ADS/JAO.ALMA\#2015.1.00389.S, ADS/JAO.ALMA\#2015.1.00405.S, ADS/JAO.ALMA\#2015.1.00587.S, ADS/JAO.ALMA\#2015.1.00820.S, ADS/JAO.ALMA\#2015.1.01012.S, ADS/JAO.ALMA\#2015.1.01120.S, ADS/JAO.ALMA\#2015.1.01225.S, ADS/JAO.ALMA\#2016.1.00177.S, ADS/JAO.ALMA\#2016.1.00329.S, ADS/JAO.ALMA\#2016.1.00852.S, ADS/JAO.ALMA\#2016.1.00948.S, ADS/JAO.ALMA\#2016.1.01172.S, ADS/JAO.ALMA\#2016.1.01265.S, ADS/JAO.ALMA\#2016.1.01269.S, ADS/JAO.ALMA\#2017.1.00025.S, ADS/JAO.ALMA\#2017.1.00496.S, ADS/JAO.ALMA\#2017.1.00601.S, ADS/JAO.ALMA\#2017.1.00629.S, ADS/JAO.ALMA\#2017.1.01093.S, ADS/JAO.ALMA\#2017.1.01727.S, ADS/JAO.ALMA\#2018.1.00541.S, ADS/JAO.ALMA\#2018.1.00558.S, ADS/JAO.ALMA\#2018.1.00940.S, ADS/JAO.ALMA\#2018.1.01852.S, ADS/JAO.ALMA\#2019.1.00260.S, ADS/JAO.ALMA\#2019.1.00597.S, ADS/JAO.ALMA\#2019.1.01136.S, ADS/JAO.ALMA\#2019.1.01757.S, ADS/JAO.ALMA\#2021.1.00094.S, ADS/JAO.ALMA\#2021.1.00602.S, ADS/JAO.ALMA\#2021.1.01089.S, ADS/JAO.ALMA\#2022.1.00482.S. ALMA is a partnership of ESO (representing its member states), NSF (USA) and NINS (Japan), together with NRC (Canada), NSTC and ASIAA (Taiwan), and KASI (Republic of Korea), in cooperation with the Republic of Chile. The Joint ALMA Observatory is operated by ESO, AUI/NRAO and NAOJ. The National Radio Astronomy Observatory is a facility of the National Science Foundation operated under cooperative agreement by Associated Universities, Inc.

The Legacy Surveys consist of three individual and complementary projects: the Dark Energy Camera Legacy Survey (DECaLS; Proposal ID \#2014B-0404; PIs: David Schlegel and Arjun Dey), the Beijing-Arizona Sky Survey (BASS; NOAO Prop. ID \#2015A-0801; PIs: Zhou Xu and Xiaohui Fan), and the Mayall z-band Legacy Survey (MzLS; Prop. ID \#2016A-0453; PI: Arjun Dey). DECaLS, BASS and MzLS together include data obtained, respectively, at the Blanco telescope, Cerro Tololo Inter-American Observatory, NSF’s NOIRLab; the Bok telescope, Steward Observatory, University of Arizona; and the Mayall telescope, Kitt Peak National Observatory, NOIRLab. Pipeline processing and analyses of the data were supported by NOIRLab and the Lawrence Berkeley National Laboratory (LBNL). The Legacy Surveys project is honored to be permitted to conduct astronomical research on Iolkam Du’ag (Kitt Peak), a mountain with particular significance to the Tohono O’odham Nation.

NOIRLab is operated by the Association of Universities for Research in Astronomy (AURA) under a cooperative agreement with the National Science Foundation. LBNL is managed by the Regents of the University of California under contract to the U.S. Department of Energy.

This project used data obtained with the Dark Energy Camera (DECam), which was constructed by the Dark Energy Survey (DES) collaboration. Funding for the DES Projects has been provided by the U.S. Department of Energy, the U.S. National Science Foundation, the Ministry of Science and Education of Spain, the Science and Technology Facilities Council of the United Kingdom, the Higher Education Funding Council for England, the National Center for Supercomputing Applications at the University of Illinois at Urbana-Champaign, the Kavli Institute of Cosmological Physics at the University of Chicago, Center for Cosmology and Astro-Particle Physics at the Ohio State University, the Mitchell Institute for Fundamental Physics and Astronomy at Texas A\&M University, Financiadora de Estudos e Projetos, Fundacao Carlos Chagas Filho de Amparo, Financiadora de Estudos e Projetos, Fundacao Carlos Chagas Filho de Amparo a Pesquisa do Estado do Rio de Janeiro, Conselho Nacional de Desenvolvimento Cientifico e Tecnologico and the Ministerio da Ciencia, Tecnologia e Inovacao, the Deutsche Forschungsgemeinschaft and the Collaborating Institutions in the Dark Energy Survey. The Collaborating Institutions are Argonne National Laboratory, the University of California at Santa Cruz, the University of Cambridge, Centro de Investigaciones Energeticas, Medioambientales y Tecnologicas-Madrid, the University of Chicago, University College London, the DES-Brazil Consortium, the University of Edinburgh, the Eidgenossische Technische Hochschule (ETH) Zurich, Fermi National Accelerator Laboratory, the University of Illinois at Urbana-Champaign, the Institut de Ciencies de l’Espai (IEEC/CSIC), the Institut de Fisica d’Altes Energies, Lawrence Berkeley National Laboratory, the Ludwig Maximilians Universitat Munchen and the associated Excellence Cluster Universe, the University of Michigan, NSF’s NOIRLab, the University of Nottingham, the Ohio State University, the University of Pennsylvania, the University of Portsmouth, SLAC National Accelerator Laboratory, Stanford University, the University of Sussex, and Texas A\&M University.

BASS is a key project of the Telescope Access Program (TAP), which has been funded by the National Astronomical Observatories of China, the Chinese Academy of Sciences (the Strategic Priority Research Program “The Emergence of Cosmological Structures” Grant \# XDB09000000), and the Special Fund for Astronomy from the Ministry of Finance. The BASS is also supported by the External Cooperation Program of Chinese Academy of Sciences (Grant \# 114A11KYSB20160057), and Chinese National Natural Science Foundation (Grant \# 12120101003, \# 11433005).

The Legacy Survey team makes use of data products from the Near-Earth Object Wide-field Infrared Survey Explorer (NEOWISE), which is a project of the Jet Propulsion Laboratory/California Institute of Technology. NEOWISE is funded by the National Aeronautics and Space Administration.

The Legacy Surveys imaging of the DESI footprint is supported by the Director, Office of Science, Office of High Energy Physics of the U.S. Department of Energy under Contract No. DE-AC02-05CH1123, by the National Energy Research Scientific Computing Center, a DOE Office of Science User Facility under the same contract; and by the U.S. National Science Foundation, Division of Astronomical Sciences under Contract No. AST-0950945 to NOAO.

This research has made use of data obtained from the 4XMM XMM-Newton serendipitous source catalogue compiled by the XMM-Newton Survey Science Centre consortium.

\section*{Data Availability}

The SALVAGE data used in this work is publicly available at: \url{https://www.canfar.net/storage/vault/list/AstroDataCitationDOI/CISTI.CANFAR/25.0077/data}.

%The LDA and random forest models trained on the image qualities tested in this work are available at www.canfar.net/citation/landing?doi=23.0031. Alternatively, new models can be trained using our synthetic SKIRT images (with and without degradation to specific image qualities) of TNG100 mergers and non-merger controls which are also available at the same online repository. The images can be degraded to any image qualities using the \texttt{RealSim} code which is available at https://github.com/cbottrell/RealSim. 

\bibliographystyle{mnras}
\bibliography{Bibliography}

\appendix
\newpage
\input{AppendixA}

\input{AppendixB}

% Don't change these lines
\bsp	% typesetting comment
\label{lastpage}
\end{document}

%% file: S1_Introduction.tex
\section{Introduction}
\label{Intro}

A variety of mechanisms have been suggested to be capable of shutting down the star formation throughout a galaxy by either removing the molecular gas (the fuel for star formation) and/or inhibiting gas from collapsing into stars \citep{Man18}. One such process is the energetic feedback from an active galactic nucleus (AGN), which refers to the process of gas accretion onto the super massive black hole at the centre of a galaxy. Although there is mounting evidence that AGN feedback plays a significant role in the shutdown of star formation from simulations \citep{Hopkins08, Hopkins09-effectofgas, Dubois13, Terrazas20, Zinger20, Mercedes-Feliz23, Frosst25} and observations \citep{Sanchez18, Piotrowska22, Bluck22, Bluck23}, direct observational support for gas depletion (and the subsequent shutdown of star formation) via AGN feedback remains mixed.

Motivated by observations of radiatively efficient and inefficient AGN accretion modes \citep[e.g.,][]{Best12}, many galaxy simulations of AGN impose their effect on the interstellar medium (ISM) through a bimodal physics prescription in which radiative and kinetic feedback heats and/or removes the surrounding gas \citep{Hopkins08, Hopkins09-effectofgas, Terrazas20, Zinger20, Frosst25}. In kinetic mode feedback, collimated jets and winds can remove gas from the galaxy, depositing it in the circumgalactic medium \citep[CGM;][]{Mercedes-Feliz23, Dong25, Sivasankaran25}. The resulting gas distribution has a central cavity that varies in radius from $\thicksim$0.1-1 kpc  depending on the black hole mass and accretion rate \citep{Torrey20}. The same jets can shock heat the CGM, which prevents gas from cooling onto the disk and replenishing the star-forming fuel \citep{Zinger20}. The radiative feedback mode can also heat gas in place, inhibiting it from cooling into stars \citep{Cielo18}. 

Observational studies of the effect of AGN feedback on gas reservoirs have led to conflicting results. In support of molecular gas depletion via AGN, AGN hosts \emph{at high-redshift} have been found to have globally depleted molecular gas reservoirs compared to inactive controls \citep{Fiore17, Kakkad2017, Circosta21, Bertola24, Molyneux25} and with fast molecular outflows \citep{Spilker25}. Furthermore, at low-redshift, AGN jets have been observed ejecting gas several kpc from the nucleus \citep{Blandford19}. However, studies of AGN hosts at low-redshift find normal or enhanced molecular gas reservoirs, compared to inactive controls \citep{Maiolino97, Scoville03, Bertram07, Saintonge17, Rosario18, Jarvis20, Shangguan20, Koss21, Salvestrini22, Yu22, Molina23, Bazzi25}. How can we reconcile the individual cases of AGN-induced molecular gas outflows at low-redshift but the lack of systematic depletion in large samples of AGN? 

There are (at least) three confounding observational constraints that are impeding our understanding of the effect of AGN feedback on molecular gas: (1) AGN strength (and its ability to couple with the ISM, e.g. inclination angle), (2) spatial scales over which the AGN can directly influence the ISM, and (3) the timescale of AGN fueling and feedback compared to the window for which an AGN can be detected by conventional methods.

Regarding point (1), one might expect that stronger AGN will be more effective at removing molecular gas. However, several studies have found that more luminous AGN host more molecular gas \citep[e.g.][]{Molina23, Salome23, Chen26}. Indeed, high AGN luminosity is also an indication of strong ongoing accretion, for which high gas surface densities are required \citep{Chen26}.

Regarding point (2), one way to resolve the discrepancy between the gas removal seen in simulations and the lack thereof in observations is to consider that AGN may only impact the molecular gas reservoir immediately surrounding the AGN, while leaving the remaining gas relatively untouched. \citet{Garcia-Burillo24} find a strong signal of depletion in the central $r<50$ pc, but only for the most luminous AGN \citep[see also][]{GB21, Elford24}. At larger scales ($r\thicksim1$ kpc), the picture is less clear, with recent works reporting systematic central gas depletion in AGN \citep{Ellison21}, a diversity of gas profiles \citep{RA22, Yu22}, or no difference from non-AGN \citep{Rosario18, Molina21}. However, to date, kpc-scale measurements of the central molecular gas in AGN have been limited to small sample sizes.

Regarding point (3), a clear connection between AGN and molecular gas depletion may be elusive due to the timescales over which AGN can be detected. For one, the AGN duty cycle itself is thought to be rather short \citep{Hickox14, Schawinski15}. Moreover, conventional AGN detection methods may only have a brief window for which the AGN can be reliably identified \citep[e.g.,][]{Blecha18}. Lastly, we do not have a reliable way of knowing if the inactive galaxies themselves are recent AGN that since shut off.

To overcome the connected observational challenges listed above, we require direct observations of the molecular gas distribution in a large sample of AGN host galaxies (and non-AGN controls). Due to the diversity in AGN strengths and the stochasticity of AGN observability, a large sample is integral to reach a broad conclusion about the effect of AGN on the ISM and identify the regimes in which AGN feedback is effective or ineffective. However, resolving maps of the molecular gas emission directly requires expensive observations from oft oversubscribed observatories such as the Atacama Large (sub-)Millimetre Array (ALMA).

The SDSS-ALMA Legacy-Value Archival Gas Exploration (SALVAGE) dataset is a heterogeneous but complete sampling of galaxies in the main galaxy sample of the Sloan Digital Sky Survey (SDSS) with resolved CO(1-0) observations in the ALMA Science Archive \citep{Wilkinson26}. Together, the optical and millimetre data provide a semi-resolved perspective (here meaning the ability to distinguish the inner and outer regions of each galaxy, independently) of the star-formation, stellar mass, and molecular gas mass of 277 galaxies at $0.02\lesssim z \lesssim 0.25$. SALVAGE is well-suited to test if AGN have an impact on the surrounding molecular gas at kpc scales. The inner region of SALVAGE galaxies is set by the 3" SDSS fibre and so probes the inner $r \lesssim 1.3$ kpc regions independently from the outer region. We can therefore use SALVAGE to test if AGN deplete molecular gas from the central 1-2 kpc region, relative to the global gas reservoir.

The paper is constructed as follows. In Section 2, we describe the SALVAGE dataset and supplementary multiwavelength archival data that will assist in our AGN selection. In Section 3, we select AGN (and inactive controls) and in Section 4, we test if they are systematically depleted in central molecular gas. We discuss our results in Section 5, and present a summary of our results in Section 6. Throughout, we assume a flat $\Lambda$CDM cosmology with $\Omega_\text{M} = 0.3$, $\Omega_\Lambda = 0.7$, and H$_0 = 70$ km s$^{-1}$ Mpc$^{-1}$.

%% file: S2_Data.tex
\section{Data}
\label{Methods}

    The SALVAGE dataset is a combination of optical spectro-photometric catalogues from the SDSS and millimitre CO(1-0) data from the ALMA science archive. The optical data provides independent measurements of star formation rate (SFR) and stellar mass (M$_\star$) from the central 1.5 arcsec and outer annulus between 1.5 arcsec and the outer extent of the galaxy. Extracted from the same apertures as the SDSS data, the ALMA CO(1-0) data provide an independent measurement of the molecular gas mass (M$_\text{mol}$) in the inner and outer regions of the galaxy. SALVAGE therefore brings together a semi-resolved perspective of the components involved in the exchange between gas and stars. 
    
    In this work, we use the data from the publicly-available SALVAGE\footnote{\url{https://www.canfar.net/storage/vault/list/AstroDataCitationDOI/CISTI.CANFAR/25.0077/data}}. For a detailed description of the data products and reduction methods, we refer readers to \citet{Wilkinson26}. However, we briefly review the relevant aspects of the data necessary for this work in following subsections.

    \subsection{Data Products from the SDSS}
    \label{sdssdata}

    The Max-Planck-Institute for Astrophysics–John Hopkins University (MPA-JHU) catalogues\footnote{\url{https://wwwmpa.mpa-garching.mpg.de/SDSS/DR7/}} provide flux measurements of up to 12 emission lines, photometric stellar mass estimates, and hybrid photometry/spectral index SFR estimates \citep{K03, Tremonti04, Brinchmann04}. While the original works are based on SDSS DR4 data, the most up-to-date catalogue is for SDSS DR7 and includes some changes from the methods discussed in \citet{K03} and \citet{Brinchmann04}. Here we summarize key details regarding the data products used in this work.

    \subsubsection{Emission line fluxes}

    Emission line flux measurements are taken directly from the public MPA-JHU raw data catalogue\footnote{\url{https://wwwmpa.mpa-garching.mpg.de/SDSS/DR7/raw\_data.html}}. Fluxes in the catalogue are corrected for foreground Galactic reddening following \citet{Odonnell94}. We correct for the internal galactic reddening of the target galaxy using a Milky Way extinction curve from \citet{Cardelli89} and an assumed intrinsic ratio of H$\alpha$/H$\beta = 2.86$.

    \subsubsection{Stellar masses}

    Stellar mass estimates are determined following similar methods as \citet{K03}, in the sense that a best-fit model spectrum from \citet{BC03} is used to estimate the mass-to-light ratio (among other properties) of each galaxy. The stellar mass is estimated by multiplying the mass-to-light ratio by the luminosity derived from the photometry. The key difference is that \citet{K03} used spectral indices from the fibre spectrum, while the updated DR7 catalogue uses the optical photometry; using optical photometry allows for consistent and separate fits to the fibre and total photometry and has no systematic offset at $M_\star>10^9M_\odot$\footnote{\url{https://wwwmpa.mpa-garching.mpg.de/SDSS/DR7/mass\_comp.html}}. Multiplying the mass-to-light ratio inferred from fibre photometry by the luminosity within the fibre gives the stellar mass within the 3" fibre aperture which we refer to as \emph{inner stellar mass} ($M_{\star\text{, inner}}$). Multiplying the mass-to-light ratio inferred from global photometry by the total luminosity gives the stellar mass of the entire galaxy which we refer to as \emph{total stellar mass} ($M_{\star\text{, total}}$). The difference between the two measurements corresponds to the mass in the outer annulus of the galaxy, which we refer to as \emph{outer stellar mass} ($M_{\star\text{, outer}}$). The stellar masses have a typical uncertainty of $\thicksim$0.1 dex.

    \subsubsection{Star formation rates}

    Star formation rates from the MPA/JHU catalogue are delivered for both star-formation within the fibre aperture and the total extent of the galaxy. Within the SDSS fibre, SFR is measured on the basis of many emission lines (specifically H$\alpha$, H$\beta$, [OIII]$\lambda5007$, [NII]$\lambda6584$, [OII]$\lambda3727$, and [SII]$\lambda6716$), with the greatest weight carried by H$\alpha$ \citep[as in ][]{Brinchmann04}. For cases where there is no H$\alpha$ emission or when there is AGN contamination, SFRs within the fibre are derived from the 4000 Å break, calibrated using the H$\alpha$ sSFR estimates of star-forming galaxies. We refer to SFRs calculated within the fibre as \emph{inner SFR} (SFR$_\text{inner}$). The \emph{outer SFR} (SFR$_\text{outer}$) is estimated by isolating the optical photometry outside the fibre and performing model fitting following the method of \citet[][]{Salim07}. The SFR measured outside the fibre is therefore independent from that measured within the fibre and is added to the SFR inside the fibre to get the \emph{total SFR} (SFR$_\text{total}$). The measurements of SFR$_\text{total}$ for the AGN hosts in SALVAGE may use a different method (i.e. the 4000 Å break) than non-AGN (i.e. H$\alpha$ or the 4000 Å break). However, SFRs are mostly used for identifying reasonable controls and we therefore do not expect the difference in SFR estimations to have a significant impact on our results. We also compute the distance of each galaxy from the star-forming main sequence, $\Delta$SFR, by subtracting SFR$_\text{total}$ from the median SFR of star forming galaxies matched in stellar mass, redshift, and environment \citep[see][for more details]{Wilkinson26}.

    \subsubsection{Edge-on classification}

    Apart from the MPA/JHU catalogue, we also use the Galaxy Zoo DESI morphology catalogue \citep{Walmsley23} which uses images from the Dark Energy Spectroscopic Instrument (DESI) Legacy Imaging Surveys \citep[DESI-LS;][]{Dey19} and classifications from the public to predict key morphological features. Specifically, we use the \texttt{disk-edge-on\_yes\_fraction}, which is the fraction of votes indicating if there is a stellar disk that is being viewed edge-on. In this work, we use a threshold of \texttt{disk-edge-on\_yes\_fraction} $ > 0.5$ to identify edge-on galaxies.

    \subsection{Data Products from the ALMA Science Archive}
    \label{stacking}

    To be included in SALVAGE, galaxies from the SDSS must have a match in the ALMA science archive within an on-sky angular tolerance of four SDSS $r$-band Petrosian radii. We found 298 SDSS galaxies with observations in the ALMA archive that passed quality assurance, spectral coverage of the redshifted CO(1-0) line, spatial resolution smaller than the SDSS central fibre (3 arcsec) and a maximum resolvable scale larger than 1.5 Petrosian radii so as to not resolve out significant flux \citep[see][]{Wilkinson26}. The ALMA data are calibrated using the restored pipeline calibrations (or the calibrated data is provided by the ALMA Help Desk when pipeline calibrations are not available) and the data are imaged using the PHANGS-ALMA reduction pipeline \citep{Leroy21-pipeline}. Using the PHANGS-ALMA pipeline allows for automated processing that is reproducible by others. After quality checks of the reduced cubes, SALVAGE consists of 277 galaxies \citep[see][]{Wilkinson26}. 

    The cleaned CO(1-0) spectral cubes allow us to measure the molecular gas mass on the same scales as the SDSS optical data products (Section \ref{sdssdata}). Previously in \citet{Wilkinson26}, we extracted ``inner,'' ``outer,'' and ``total'' spectra from the cleaned cube by blindly co-adding the spectra that fall within apertures equal to the SDSS data: for the inner spectrum, we include pixels within the central 3"-diameter aperture; for the outer spectrum, we include pixels that fall within an annulus with a radial extent from 1.5" to the 4 scale lengths of the best fit SDSS photometric model; for the total spectrum, we include all pixels within a radius equal to 4 scale lengths of the best fit SDSS photometric model. However, many galaxies have non-detections in the inner, outer, or total regions.
    
    In this work, we present an improved catalogue of CO line luminosities measured from the ALMA cubes that incorporates stacking procedures to reduce the number of non-detections. To reduce the number of cases of non-detections in the inner or outer regions, we conduct a spectral stacking routine on the spectral cube. The spectral stacking method described in this section differs from the spatial summing of spectra conducted in \citet{Wilkinson26} as the new stacking method uses a 2D velocity profile to shift each spectrum to the same rest velocity before co-adding, ultimately improving S/N. The spectral stacking routine is done using the public Python code \texttt{stackarator} \citep{Davis25}\footnote{\url{https://github.com/TimothyADavis/stackarator}}. \texttt{stackarator} takes as input the spectral cube and a moment 1 map and returns a stacked spectrum and a new (lower) RMS noise value. The details for the inputs passed to \texttt{stackarator} are as follows:

\textbf{Imaged ALMA Cube:} In this work, we use the same imaged ALMA cubes as described in \citet{Wilkinson26}. Although \texttt{stackarator} can define regions within a galaxy to stack, we create our own masked cubes before passing to \texttt{stackarator} to ensure the masking process is identical to the methods described in \citet{Wilkinson26}.
 
\begin{figure*}
        \centering
        \includegraphics[width=1\linewidth]{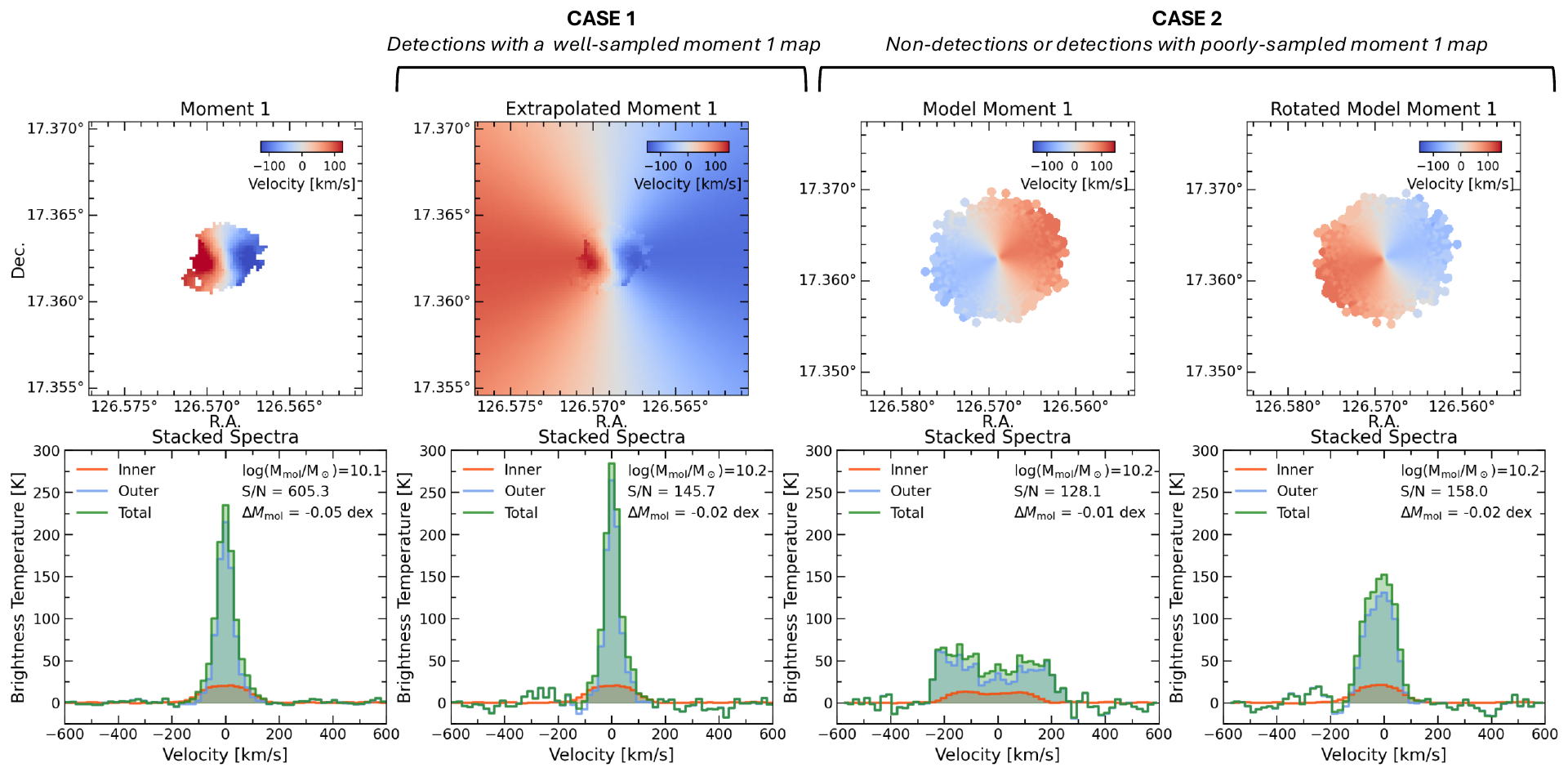}
        \caption[A summary of the spectral stacking input data and resulting spectra.]{A summary of the spectral stacking input data and resulting spectra for SDSS ObjID 587741490891325518. The top row shows the original moment 1 map (left) and the three options for moment maps used as input to \texttt{stackarator}. Proceeding from left to right, the options are the moment 1 map extrapolated to non-detected spaxels using a best-fit exponential disk model, the model moment 1 map generated using only the morphology data from the optical imaging, and the same model moment 1 map but rotated by 180 degrees. Along the bottom row are the resulting stacked spectra for the inner (red), outer (blue), and total (green) regions. In the top right corner of the bottom panels is the total molecular gas mass reported for this galaxy, along with its S/N and difference from the molecular gas mass measured using co-add methods in \citet{Wilkinson26}. When available (like this case), the extrapolated moment 1 map measures reasonable molecular gas mass, with a strong S/N. In cases where the moment 1 map is not available, the model moment 1 map with the better S/N (in this case, the rotated model) produces a consistent molecular gas mass measurement, albeit with a larger line width indicating small errors in the spaxel velocity estimates.}
        \label{stackedspectra}
    \end{figure*}

\textbf{Moment 1 Map:} The moment 1 map passed to \texttt{stackarator} depends on whether the ALMA observation is deep enough to have a reasonably well-sampled moment 1 map or not and we present a summary of the different input moment 1 maps in Figure \ref{stackedspectra}. Specifically, we define a well-sampled moment map as having at least 50 detected spaxels. Of the 277 galaxies in SALVAGE, there are 200 with a well-sampled moment map. In these cases, which we refer to as CASE 1, we use the moment 1 map output by the PHANGS-ALMA reduction pipeline and interpolate/extrapolate non-detected spaxels assuming an exponential rotating disk model fit to the detected moment 1 map spaxels. The best fit process uses the \texttt{scipy} \texttt{curve\_fit} function with the following values used as the initial values of the fit parameters: 

\begin{itemize}
    \item the system velocity is initialized at the median velocity of the moment map, 
    \item the initial rotation velocity is initialized at 300 km/s,
    \item the inclination is initialized at the value from the fits to the SDSS $r$-band images, and
    \item the position angle is initialized at the value from the fits to the SDSS $r$-band images.
\end{itemize} 

For the remaining 77 galaxies with fewer than 50 detected spaxels in the moment 1 map (CASE 2), we compute a model moment 1 map using the public Python package \texttt{KinMS}\footnote{\url{https://github.com/TimothyADavis/KinMSpy/tree/master}}. The function \texttt{expdisk} in \texttt{KinMS} creates a synthetic spectral cube of a rotating exponential disk rotating taking into account observational effects such as beam smearing, which we adapt to incorporate the data available. Specifically, we modify the \texttt{expdisk} function to create the exponential disk model using the optical radius measurement, ellipticity (converted to inclination), and position angle measured from the SDSS optical morphology. However, using the optical data to infer the disk structure does not inform us as to which way the galaxy is rotating. Therefore, we create two model moment 1 maps (with opposite rotations) and take the one that delivers the highest S/N stacked spectra. 

    We convert the observed line fluxes ($S_\text{CO}$) to CO(1-0) integrated line luminosities ($L'_\text{CO}$), following equation (3) from \citet{Solomon05}:

    \begin{equation}
        L'_\text{CO} = 3.25\times10^7S_\text{CO}\nu^{-2}D_L^2(1+z)^{-3},
        \label{LCOeq}
    \end{equation}

    \noindent where $\nu$ is the expected frequency of the line according to the SDSS spectroscopic redshift ($z$) in GHz and $D_L$ is the luminosity distance in Mpc.

    The CO(1-0) integrated line luminosity is converted to a molecular gas mass estimate, $M_\text{mol}$, using a constant conversion factor of $\alpha_\text{CO} = 4.35$ M$_\odot$ (K km s$^{-1}$ pc$^{2}$)$^{-1}$ \citep{Schinnerer24}:

    \begin{equation}
        M_\text{mol} = \alpha_\text{CO}L'_\text{CO}.
        \label{MH2eq}
    \end{equation}

    \noindent Our end result is molecular gas masses extracted from the inner, outer, and total regions that match the spatial scales of the SDSS optical data products, which we call M$_\text{mol, inner}$, M$_\text{mol, outer}$, and M$_\text{mol, total}$.

    A summary of the moment maps that are passed to \texttt{stackarator} and the resulting stacked spectra are shown in Figure \ref{stackedspectra}. The top row of Figure \ref{stackedspectra} shows the moment maps passed as input to \texttt{stackarator} for an example galaxy (SDSS ObjID 587741490891325518) that has a robust moment 1 map (i.e. more than 50 detected spaxels in the original moment map). First, we will compare the performance between the regular moment 1 map and the extrapolated moment 1 map using a best fit exponential model (i.e. CASE 1). On the left, the moment 1 map looks robust, reporting values only for well-detected spaxels. However, it does have some spaxels within the disk that are missing data, as well as an outer edge that does not reach the full extent of the outer aperture. As a result, when we measure the total molecular gas (reported in the upper right corner of the bottom panel), we get a value that is 0.05 dex lower than the value we measured from co-adding the spectra from the entire outer region \citep[we report the difference between the M$_\text{mol}$ measured here and the measurement from ][as $\Delta$M$_\text{mol}$]{Wilkinson26}. However, in the second column, the extrapolated moment 1 map allows all the spaxels to be included in the stack, using the information provided by the original moment 1 map as best as possible. The final reported molecular gas mass has a S/N that is worse than if we were to use only detected spaxels, but a M$_\text{mol}$ measurement that is in better agreement with the co-added molecular gas mass from \citet{Wilkinson26} (i.e. $\Delta$M$_\text{mol} = -0.02$ dex).  

    Next, we will compare the performance between the two model moment 1 maps made using only the optical image information (i.e. CASE 2). In the second column from the right of Figure \ref{stackedspectra}, the moment 1 map is oriented in the wrong rotation direction. The resulting stacked spectra shown in the bottom row show two distinct peaks instead of one, indicating that they are not being shifted to the correct rest velocity. However, by simply rotating the moment 1 map and trying again (rightmost column), the spectra are aligned into a single peak, with better S/N than in the previous case. The rotated model moment 1 results in a spectrum with a larger line width than spectra created using the extrapolated moment 1 map, implying a mismatch between the true velocity profile and the model moment 1 map. Therefore, we always use the molecular gas mass measurement from the extrapolated method (CASE 1) when available, as it uses the CO velocity information to inform the stacking. However, the molecular gas mass measured using the rotated model moment 1 map is still in good agreement with the extrapolated moment 1 map method and with the co-added measurements in \citet{Wilkinson26}, demonstrating that it is a reliable way of stacking non-detections.

    As a final consistency check, we compare the total molecular gas masses measured using the spectral stacking method to the original spatial co-add method in \citet{Wilkinson26} and to literature values from the xCOLD GASS \citep{Saintonge17} and MASCOT \citep{Wylezalek22} surveys. The median difference between the spectral stacking measurements and spatial co-add method is $-0.013$ dex, with a scatter of $0.13$ dex. The median difference between the spectral stack methods and the values derived from xCOLD GASS (12 galaxies in common) and MASCOT (5 galaxies in common) is 0.026 dex, with a scatter of 0.06 dex. Although there is significant scatter between the measurements, there is no systematic offset introduced by the new method. Using a quality threshold of S/N$>10$ (which we find necessary given that the stacked spectra can have very low noise values even for non-detections), we recover 17 additional galaxies that were not detected in the original catalogue at a S/N$>5$.

    \subsection{Data Products that Combine SDSS and ALMA Data}

    In this work, we take the inner, outer, and total SFR, M$_\star$, and M$_\text{mol}$ from SALVAGE and produce several higher order products intended to probe the molecular gas distribution in the central regions of AGN. To this end, we first compute the central molecular gas mass surface density, $\Sigma_\text{mol, inner}$, by dividing by the surface area of the inner region defined by the SDSS fibre:

    \begin{equation}
        \Sigma_\text{mol, inner} = \frac{\text{M}_\text{mol, inner}}{\pi r_\text{inner}^2}, 
    \end{equation}

    \noindent where $r_\text{inner}$ is the 1.5" radius of the SDSS fibre converted to a physical distance in units of kpc. %We do not compute the outer surface density since the quality of the match between the outer radius and the ``edge'' of the stellar distribution can depend on stellar morphology, introducing noise. 
    Similarly, we also compute the gas fraction of the inner region as:

    \begin{equation}
        \label{fgasinner}
        f_\text{gas, inner} = \frac{\text{M}_\text{mol, inner}}{\text{M}_{\star\text{, inner}}}.
    \end{equation}

    Lastly, we compute a self-referential diagnostic of central gas depletion by computing the ratio of the inner and outer gas fractions, which we denote as $\nabla f_\text{gas}$:

    \begin{equation}
        \nabla f_\text{gas} = \frac{f_\text{gas, inner}}{f_\text{gas, outer}}, 
    \end{equation}

    where $f_\text{gas, outer}$ is computed the same as in Equation \ref{fgasinner}, but for M$_\text{mol, outer}$ and M$_{\star\text{, outer}}$.

    \subsection{Supplementary Multiwavelength Archival Data}
\label{supplementaryarchive}

    AGN can be detected at many different wavelengths. Although AGN detection methods may overlap, there are many instances where an AGN may be detected at one wavelength and not in another \citep[e.g.,][]{Padovani2017, Bickley24-xrayagn, Alban24}. To enable AGN detection in SALVAGE that is as complete as possible, we supplement SALVAGE with archival data from the mid-IR and X-ray. 

    \subsubsection{Mid-IR WISE Photometry}

    The Wide-field Infrared Survey Explorer (WISE) space telescope has four bands in the mid-IR, referred to as W1, W2, W3, and W4 centered on 3.4, 4.6, 12, and 22 microns, respectively. In this work, we use the unWISE forced photometry catalogue, which applies aperture photometry on the positions of SDSS galaxies \citep{Lang16}. This ensures complete WISE coverage for the galaxies in our sample. 

    \subsubsection{X-Ray Luminosity Catalogues}
    \label{xraydata}

    The extended ROentgen Survey with an Imaging Telescope Array (eROSITA) has released its first data release of the eROSITA All Sky Survey (eRASS1). The German consortium half of the sky (Galactic longitudes between $180^\circ < l < 360^\circ$) is fully public. We match SALVAGE targets to the eRASS1 source catalogue\footnote{\url{https://erosita.mpe.mpg.de/dr1/AllSkySurveyData_dr1/Catalogues\_dr1/MerloniA_DR1/eRASS1\_Main.html}} with a matching tolerance equal to the resolution of eROSITA (20 arcsec). Despite $\thicksim70$\% of SALVAGE falling within the coverage of the public eRASS1 data release, only 7 SALVAGE galaxies are matched to the source catalogue. We combine the flux from the 2-5 keV and 5-8 keV bands and convert from 2-8 keV to 2-10 keV assuming an SED slope with $\Gamma=1.9$. The 2-10 keV flux ($F_\text{X, 2-10 keV}$, in units of erg s$^{-1}$ cm$^{2}$) is converted to a luminosity with the following equation:

    \begin{equation}
    \label{luminosity}
        L_\text{X, 2-10 keV}= 4\pi D_L^2F_\text{X, 2-10 keV}
    \end{equation}

    \noindent where $D_L$ is the luminosity distance in units of cm. The all-sky survey design of eROSITA aims to observe the entire sky down to a sensitivity of $\thicksim5\times 10^{-14}$ erg s$^{-1}$ cm$^{2}$ in the 2-8 keV hard X-ray band \citep{Predehl21}. Given the redshifts of SALVAGE targets, this translates to an upper limit on $L_\text{X, 2-10 keV}$ of $\thicksim 10^{41}$ erg s$^{-1}$. Therefore, a non-detection in eROSITA does not exclude the possibility of a low-luminosity X-ray AGN being present.
    
    %I choose not to correct for internal dust correction and later find that it would not affect the AGN selection (see Section \ref{xray}). 

    Chandra is a space-based observatory launched in 1999 that is available for targeted X-ray observations. All observations become public after a 1-year proprietary period and are stored on the Chandra archive. Contrary to the case for eROSITA, these observations are highly heterogeneous. The Chandra Source Catalogue\footnote{\url{https://cxc.cfa.harvard.edu/csc2.1/}} version 2.1 (CSC 2.1) was released in October 2024 and includes all observations made public prior to the end of 2021. We query the CSC 2.1 for X-ray sources within three arcseconds of SALVAGE targets (i.e. the size of the SDSS fibre) with \texttt{PyVO} and find three matches. Increasing the matching tolerance up to 20" did not increase the number of matched X-ray sources. We take the flux from the 2-7 keV band and convert it to a 2-10 keV luminosity assuming a spectral slope of $\Gamma=1.9$ and Equation \ref{luminosity}.

    XMM-Newton is a space-based observatory that was also launched in 1999. Observations from XMM-Newton have been collected and homogenized into the 4XMM Serendipitous Source Catalogue \citep{Webb20}. We search this catalogue\footnote{\url{http://xmmssc.irap.omp.eu/Catalogue/4XMM-DR14/4XMM_DR14.html}} for X-ray sources within 15" of SALVAGE targets \citep[following the 15" matching tolerance with SDSS employed in][]{Webb20} and find 21 matches. For each source, we combine the flux from the 2-4.5 keV and the 4.5-12 keV and convert it to a 2-10 keV luminosity assuming a spectral slope of $\Gamma=1.9$ and Equation \ref{luminosity}.

    We also crossmatch SALVAGE with the \emph{Swift}/BAT AGN catalogue \citep{Baumgartner13}, NuSTAR Serendipitous Survey \citep{Greenwell24}, and the ROSAT BSC/FSC catalogue \citep{Paronynan21} and find no matches. 

    \subsubsection{Radio AGN Catalogues}

    We search for radio AGN in SALVAGE using several catalogues. \citet{Best12} compiled a catalogue of high- and low-excitation radio galaxies (HERGs and LERGs) by crossmatching SDSS DR7 targets with the NRAO VLA Sky Survey \citep[NVSS;][]{Condon98} and the Faint Images of the Radio Sky at Twenty centimetres (FIRST) survey \citep{Becker95}. From the \citet{Best12} catalogue, we find two LERGs in SALVAGE. We also cross-match SALVAGE with the FIRST radio AGN catalogue from \citet{Lofthouse18} and find 3 matches within 5 arcseconds. Lastly, we cross-match SALVAGE with the radio AGN catalogue produced by \citet{Hardcastle25} using the LOFAR Two-Metre Sky Survey \citep[LoTSS;][]{Shimwell17} 144 MHz emission. We find one match between LoTSS AGN and SALVAGE. However, all six of the radio AGN identified in SALVAGE are either too high redshift or too inclined to be included in this work. 

%% file: S3_Sample.tex
\section{AGN Sample Selection}
\label{ch4-sample}

    The parent sample of this study is constructed from the publicly available SALVAGE catalogue containing 277 galaxies. To ensure the inner values taken from SALVAGE are truly probing the central regions, the GalaxyZoo edge-on probability (see Section \ref{sdssdata}) is required to be less than 0.5, limiting the sample to 242 galaxies. Since the distribution of SALVAGE targets in redshift \citep[$0.02\lesssim z\lesssim0.25$; see][]{Wilkinson26} allows for a large range of physical sizes of the inner region, we limit the analysis to galaxies with $r_\text{inner}< 2$ kpc, as defined by the 3" central fibre and the redshift of the system. This corresponds to a maximum redshift of $z=0.0698$ and limits the sample further to 199 galaxies. Of these 199 galaxies in SALVAGE that have $r_\text{inner}<2$ kpc and are not viewed edge-on, 159 of them have stacked global gas detections with S/N $> 10$ and at least an inner \emph{or} outer detection with S/N $> 10$. The parent sample used in this work therefore consists of 159 galaxies in SALVAGE that have a strong CO detection, $r_\text{inner}<2$ kpc, and are not viewed edge-on.
    
    We note that, even after improving our detection rate with spectral stacking, 23 of the 159 targets in this work are still missing an inner or outer CO detection with S/N $> 10$. These partial non-detections that could not be recovered from spectral stacking require mindful interpretation in the subsequent analysis. We incorporate upper limits in our assessment when feasible. 
     
    In this Section, we describe the selection of AGN from optical emission lines (Sections \ref{bpt} and \ref{whan}), mid-IR broadband colour (Section \ref{wise}), and X-ray luminosity (Section \ref{xray}). \emph{For a galaxy to be considered an AGN host, we require that it meets at least one of the selection criteria outlined in this section.} We then identify non-AGN controls in Section \ref{controls}. When reporting the number of galaxies identified as AGN and controls in this section, we are only considering those that meet our parent sample criteria (i.e. $r_\text{inner}<2$ kpc, not viewed edge-on, global M$_\text{mol}$ detections with S/N $> 10$ and either an inner or outer detection).

    \subsection{Optical Line Ratios: BPT Diagram}
    \label{bpt}

    AGN can be selected on the basis of optical emission line ratios that trace the strength of the ionizing radiation affecting the interstellar medium (ISM). The ``BPT diagram'' \citep{Baldwin81} uses the line ratios [NII]$\lambda$6584/H$\alpha$ and [OIII]$\lambda$5007/H$\beta$ to distinguish AGN-like emission from line emission consistent with star-formation. In this work, we use the BPT AGN classification criteria from \citet{K03} and \citet{Kewley01}. The \citet{K03} criterion is a more inclusive threshold that includes both ``composite'' AGN (ionization consistent with star formation and AGN) and ``pure'' AGN (ionization dominated by AGN), while the \citet{Kewley01} criterion aims to select only pure AGN. Specifically, to be selected as an AGN by \citet{K03}, the emission lines from the galaxy must satisfy the following condition:

    \begin{equation}
        \log\left(\text{[OIII]}/\text{H}\beta\right) > 0.61 / \log\left(\text{[NII]}/\text{H}\alpha) - 0.05\right) + 1.3.
    \end{equation}
    
    \noindent Similarly, to be considered a \citep{Kewley01} AGN, the emission lines from the galaxy must satisfy the following condition:

    \begin{equation}
        \log\left(\text{[OIII]}/\text{H}\beta\right) > 0.61 / \log\left(\text{[NII]}/\text{H}\alpha) - 0.47\right) + 1.19.
    \end{equation}

    In this work, we will refer to galaxies whose emission lines satisfy the \citet{K03} AGN criterion, but not the \citet{Kewley01} criterion as composite AGN. Meanwhile, galaxies that satisfy the \citep{Kewley01} AGN criteria will be referred to as K01 AGN. Furthermore, in order to be classified as an AGN by either of these methods, the following criteria must also be met:

    \begin{itemize}
        \item Must have S/N $>3$ for the [NII], H$\alpha$, [OIII], and H$\beta$ emission lines to ensure a reliable classification.
        \medskip
        \item Must have EW(H$\alpha$)$<-3$ Å.\footnote{We subscribe to the convention that the equivalent width of emission lines are given negative values.} \citet{Belfiore22} demonstrate that hard ionization from hot low-mass evolved stars (HOLMES) can lead to larger values of $\text{[OIII]}/\text{H}\beta$, particularly in the central regions of galaxies (where the SDSS fibre is located). However, the enhanced $\text{[OIII]}/\text{H}\beta$ induced by HOLMES are rarely associated with regions with EW(H$\alpha$)$<-3$ Å \citep{Stasinska06, CidFernandes11}.
    \end{itemize}
    
    In SALVAGE, there are 12 K01 AGN and 36 composite AGN with a CO detection, $r_\text{inner} < 2$ kpc and not viewed edge-on. 

    \subsection{Optical Line Ratios: WHaN Diagram}
    \label{whan}

    While the BPT diagram uses four emission lines to classify galaxies as AGN, the EW(H$\alpha$)-H$\alpha$/[NII] (WHaN) diagram introduced in \citet{CidFernandes11} selects AGN under similar principles, but using only two emission lines: [NII] and H$\alpha$. This allows galaxies that may not have sufficient S/N in [OII] and H$\beta$ to still be classified as AGN. Recent work by \citet{Sanchez25} has shown that the WHaN diagram is more effective at separating star-forming, retired and AGN ionized galaxies than the traditional BPT diagram.
    
    Following \citet{CidFernandes11}, a S/N $> 10$ cut is imposed on the g-band spectral flux of the SDSS spectra (i.e \texttt{sn0} in the SDSS catalogues). The WHaN diagram separates star-forming galaxies, strong AGN, weak AGN, and retired galaxies according to the following criteria:

    \begin{itemize}
        \item A star-forming galaxy on the WHaN diagram has log([NII]/H$\alpha$)$<-0.4$ and EW(H$\alpha$)$<-3$Å.
        \medskip

        \item A strong AGN (sAGN) on the WHaN diagram has log([NII]/H$\alpha$)$>-0.4$ and EW(H$\alpha$)$<-6$Å.
        \medskip

        \item A weak AGN (wAGN) on the WHaN diagram has log([NII]/H$\alpha$)$>-0.4$ and $-6$Å$<$EW(H$\alpha$)$<-3$Å.
        \medskip

        \item A retired galaxy on the WHAN diagram has log([NII]/H$\alpha$)$>-0.4$ and EW(H$\alpha$)$>-3$Å
        \medskip
    \end{itemize}
    
    In SALVAGE, there are 61 sAGN and 15 wAGN with a CO detection, $r_\text{inner} < $2 kpc and not viewed edge-on. There is significant overlap between the WHaN diagram AGN classification and the BPT diagram. Of the 61 sAGN, 35 are also identified as AGN by the BPT diagram (27 composite, 8 K01). Of the 15 wAGN, 12 are identified as AGN by the BPT diagram (8 composite, 4 K01).

    \subsection{Mid-IR Colour}
    \label{wise}

    Ultraviolet emission from the AGN accretion disk can heat the circumnuclear or galaxy-scale dust that might otherwise obscure the optical emission lines \citep{Hickox18}. Therefore, we also identify AGN from the slope of their spectral energy distribution driven by the dust emission in the mid-IR. For this, we use WISE photometry bands W1 and W2 centred at 3.4 and 4.6 microns, respectively. While \citet{Stern12} proposed a colour cut of W1$-$W2$>0.8$ to identify AGN, theoretical work from \citet{Blecha18} has shown that W1$-$W2$>0.5$ can reliably identify low-redshift mid-IR AGN with higher completeness. To maximize completeness in the AGN sample, we identify mid-IR AGN using the \citet{Blecha18} proposed threshold of W1$-$W2$>0.5$. In SALVAGE, there are 3 mid-IR AGN with a CO detection, $r_\text{inner} < $2 kpc and not viewed edge-on. Two of the three mid-IR AGN are identified as sAGN in the WHaN diagram, but one is a unique AGN not identified by any other method in this work.
    
    \subsection{X-ray Luminosity}
    \label{xray}

    \begin{figure}
    \label{EulerDiagram}
        \centering
        \includegraphics[width=1\linewidth]{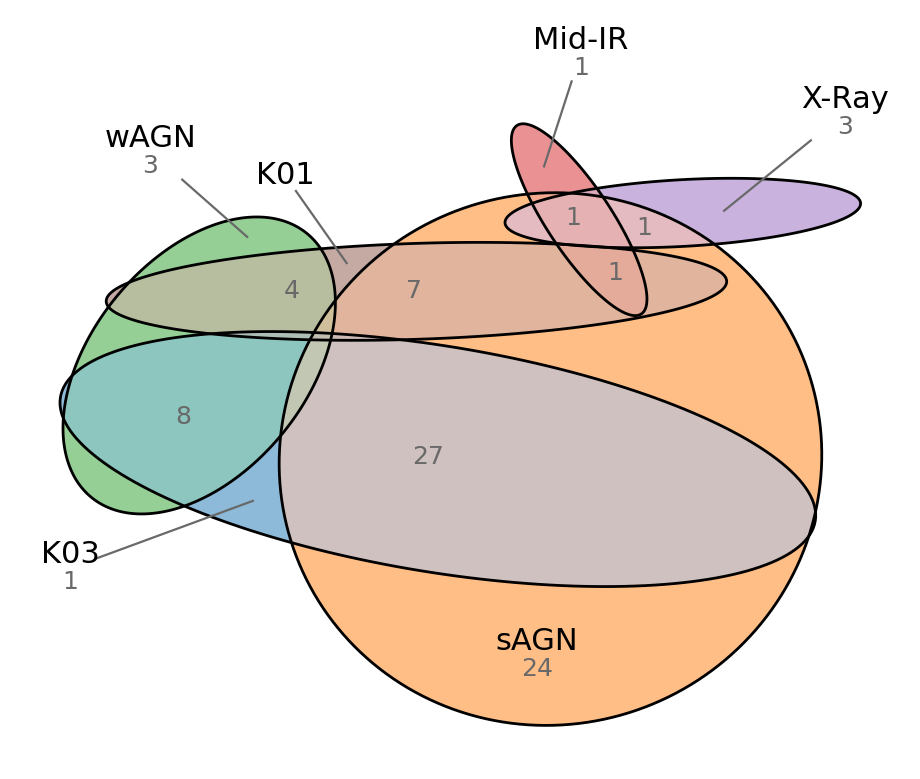}
        \caption{An Euler diagram representing the relative size and overlap of the six AGN classes used in this work. The area of each ellipse is proportional to number of targets classified by that AGN class, and the area of overlapping regions is approximately proportional to the overlap between each category (gray numbers). Regions with no quantity label (gray number) can be assumed to be zero.}
    \end{figure}

    In the hot corona around the AGN, UV photons from the accretion disk are excited to X-ray energies via inverse Compton scattering interaction with relativistic electrons \citep{Haardt91, Stern95}. As a result, high X-ray luminosities are a reliable AGN indicator. However, X-rays can also be emitted from other sources such as X-ray binaries and hot plasmas associated with star forming regions that correlate with SFR \citep{Ranalli03, Lehmer10, Mineo12}. Therefore, we follow the method of \citet{Agostino19} in which the SFR-X-ray relation established in \citet{Ranalli03} is rearranged into an X-ray AGN selection criterion. In order for a galaxy with an X-ray detection to be considered an AGN, it must have an X-ray luminosity in excess of that expected given its SFR:

    \begin{equation}
    \label{xrayagncriterion}
        \log (1.33  \times L_{X, \text{ 2-10 keV}} \times 10^{-40} \text{ erg s}^{-1}) - 0.6 > \log(\text{SFR}_\text{total}),
    \end{equation}

    \noindent where a factor of 1.5 is included in the factor of 1.33 to adjust between the Salpeter IMF \citep{Salpeter1955} assumed in \citet{Ranalli03} and the Kroupa IMF \citep{Kroupa01} assumed in this work \citep{B04}. We include the 0.6 dex offset following \citet{Agostino19} to ensure that the X-ray luminosity is 2$\sigma$ higher than the expected luminosity given the \citet{Ranalli03} $L_X$-SFR relation.
    
    Using the observed 2-10 keV X-ray luminosities from the archival search (see Section \ref{xraydata}), we find that there are eight SALVAGE targets with X-ray detections and global CO detections with $r_\text{inner} < 2$ kpc and are not viewed edge-on (two from eROSITA and six from XMM-Newton). Of these eight X-ray detections, five satisfy the \citet{Agostino19} X-ray AGN criteria outlined in Equation \ref{xrayagncriterion}. 
    
    Of the five X-ray AGN, two are identified as a WHaN sAGN and one is detected as a mid-IR AGN. The two X-ray AGN that are not detected by any other AGN criteria were targeted observations by XMM-Newton. We remind the reader that SALVAGE does not have complete or homogeneous coverage at X-ray wavelengths. Although 70\% of SALVAGE is within the public eRASS1 footprint, the observations are not deep enough to rule out X-ray AGN as all upper limits on $L_\text{X}$ from eROSITA are above the expected SFR-X-ray relation. We therefore cannot rule out that the non-X-ray AGN in SALVAGE are in fact X-ray AGN. We include X-ray AGN based on the data available for completeness, but acknowledge that the X-ray coverage is not complete.

    To demonstrate the relative sample size and overlap between the different AGN selection methods described in this section, we present an Euler diagram in Figure \ref{EulerDiagram}. The area of each ellipse is proportional to the size of the AGN sample selected using the labeled method \citep[see][]{Larsson18}. Gray numbers represent the number of SALVAGE galaxies in the overlapping regions between different AGN selections.
    
    \subsection{Non-AGN Controls}
    \label{controls}

    In total, 81 SALVAGE targets are identified as an AGN host by at least one of the AGN identification methods described above. To assess the molecular gas properties of AGN hosts, we will compare them to non-AGN controls. There are 78 galaxies that are not identified as AGN by one of the methods described in this section which makeup the non-AGN control pool.

    \begin{figure*}
        \centering
        \includegraphics[width=1\linewidth]{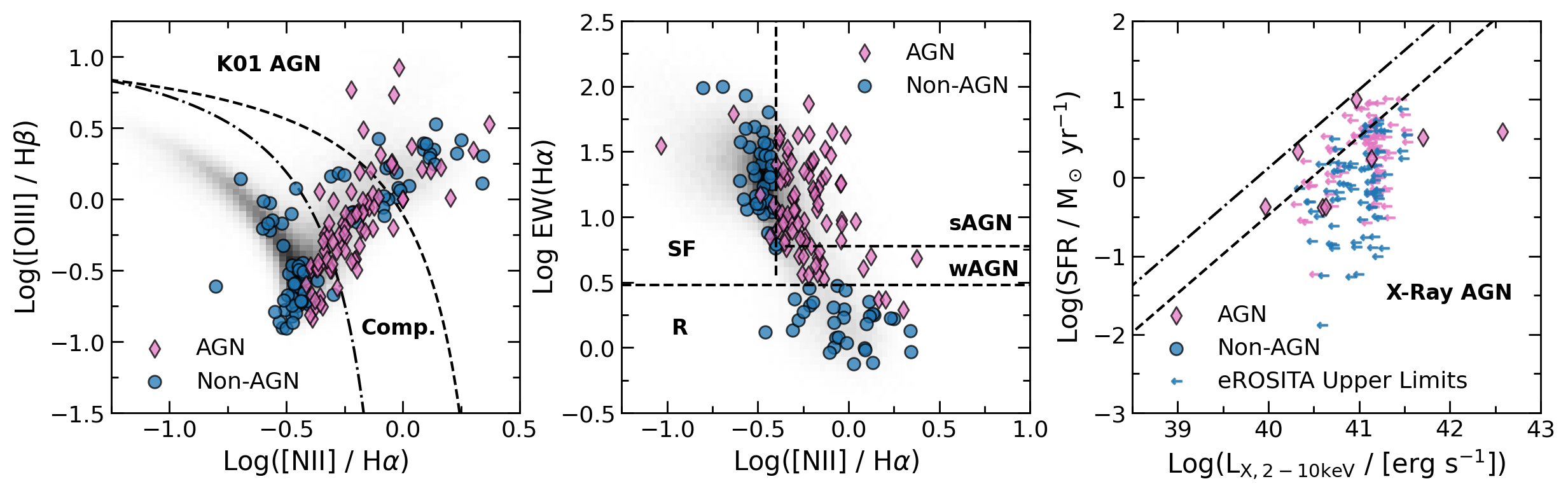}
        \caption[The positions of SALVAGE AGN and non-AGN on the BPT diagram, WHAN diagram]{\emph{Left:} BPT diagram of SALVAGE AGN (pink diamonds) and non-AGN (blue circles) that meet the parent sample selection (e.g., CO detections and not viewed edge-on) and \emph{any} of the AGN selection criteria, compared with all of SDSS (grey shading). The black dot-dash line represents the \citet{K03} AGN/SF demarcation and the black dashed line represents the \citet{Kewley01} AGN criterion. \emph{Centre:} WHaN diagram of SALVAGE AGN and non-AGN with the \citet{CidFernandes11} star forming, sAGN, wAGN, and retired classifications represented by black dashed lines. \emph{Right:} SFR-X-ray relation for the SALVAGE sample. The black dot-dashed line represents the \citet{Ranalli03} SFR-X-ray relation and the black dashed line represents 0.6 dex higher X-ray luminosity than expected from star formation. Pink diamonds show SALVAGE AGN and no non-AGN are detected in X-ray. Arrows represent AGN and non-AGN with upper limits from eROSITA. $\thicksim30$\% of SALVAGE does not have any X-ray information and is not included in this plot.}
        \label{agnsample}
    \end{figure*}

    In Figure \ref{agnsample}, we present three AGN diagnostic diagrams used in this work. In each panel, we show the positions of galaxies in SALVAGE that meet \emph{any} of the AGN criteria as pink diamonds and those that meet none of the AGN (i.e., non-AGN) as blue circles. In the left panel, we present the AGN and non-AGN targets on the BPT diagram, as well as all SDSS targets ($z<0.1$) with sufficient emission line S/N shown as a shaded 2D histogram in the background. The \citet{Kewley01} AGN criterion is shown as a black dashed line and the \citet{K03} AGN criterion is shown as a black dot-dashed line. Since the AGN sample presented here includes AGN that meet any of the AGN selection methods (i.e. including the WHaN, mid-IR and X-ray AGN), there are some AGN below the \citet{K03} AGN line. Non-AGN found above the AGN thresholds do not meet the required H$\alpha$ equivalent width. Galaxies without S/N $>3$ in all four emission lines are not present on this diagram. In the centre panel, we show the AGN and non-AGN positions on the WHaN diagram. The criteria for star-forming (SF), retired (R), sAGN, and wAGN are shown as black dashed lines. In the right panel, we show the SFR-X-ray relation from \citet{Ranalli03} as a black dot-dashed line and the X-ray AGN criterion, 0.6 dex below the relation, as a black dashed line. AGN are shown as pink diamonds and there are no X-ray-detected galaxies in the non-AGN control pool. Galaxies with eROSITA coverage but no X-ray detection are shown plotted as arrows representing the upper limit on their X-ray luminosity. 

    \begin{figure}
        \centering
        \includegraphics[width=1\linewidth]{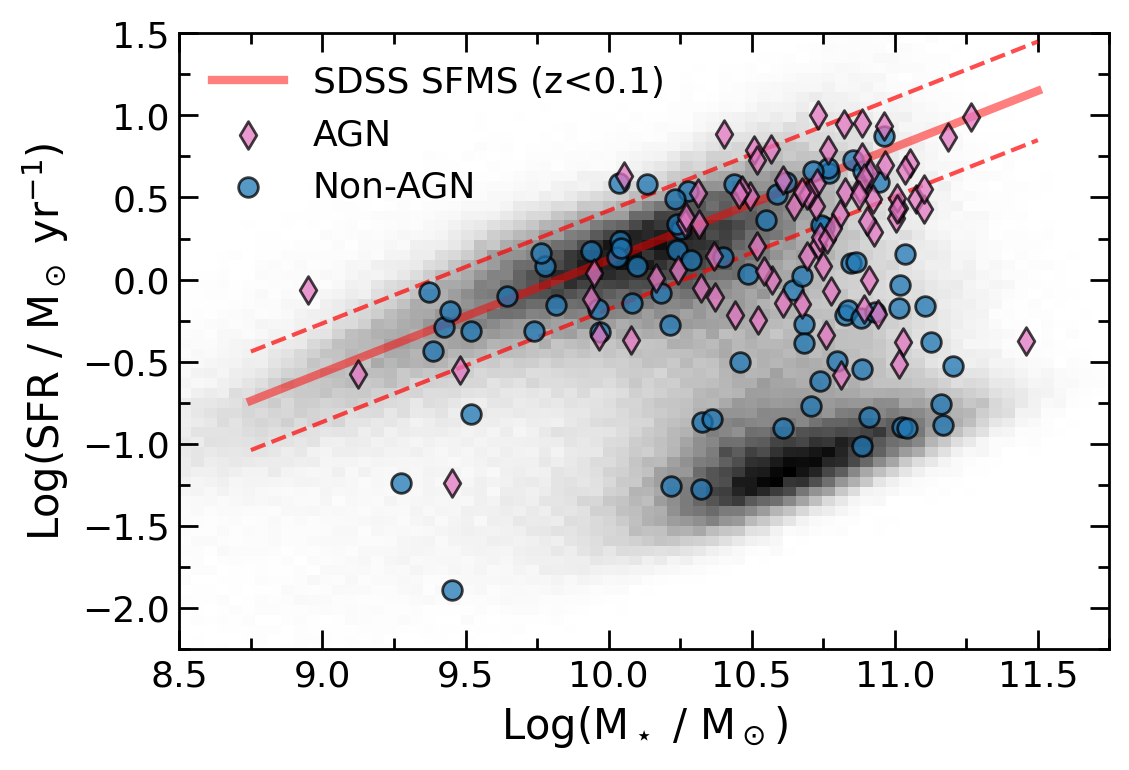}
        \caption[The positions of SALVAGE AGN and non-AGN on the SFMS.]{The positions of SALVAGE AGN (pink diamonds) and non-AGN (blue circles) on the SFMS before control matching is conducted. All of SDSS DR7 ($z<0.1$) is shown as a black 2D histogram in the background for context. The solid red line shows a linear fit to all star forming galaxies as classified by the \citet{Kauffmann03-agn} criterion on the BPT diagram to guide the eye and the dashed red lines show $\pm0.3$ dex above and below the SFMS. Both the AGN and non-AGN samples in SALVAGE tend to have high stellar masses as they are easier to observe with ALMA \citep[see][for a full discussion]{Wilkinson26}. However, the AGN span a range of stellar masses and SFRs with many similar non-AGN controls available.}
        \label{agnsfms}
    \end{figure}
    
    To contextualize the AGN and non-AGN samples in SALVAGE, we show their positions relative to the SDSS SFMS in Figure \ref{agnsfms}. To guide the eye, we conducted a linear fit to all star-forming galaxies in SDSS DR7 at $z<0.1$, as defined by the \citet{Kauffmann03-agn} criterion on the BPT diagram, shown as a solid red line. Red dashed lines represent $\pm0.3$ dex above and below the SFMS. In general, both the AGN and non-AGN in SALVAGE tend to be star forming and higher mass than typical low-$z$ galaxies in SDSS (shown as the black 2D histogram). The positions of the AGN and non-AGN relative to the SFMS reflect the overall bias of the SALVAGE sample, which preferentially includes high mass and star-forming targets as they are more likely to achieve molecular gas detections with ALMA and thus preferred by ALMA users \citep[see][]{Wilkinson26}. The AGN span a range of SFRs and stellar masses. However, there are many AGN that do not have a nearby non-AGN, but only at the high and low stellar mass extrema. Considering that the AGN and non-AGN samples share similar locations in Figure \ref{agnsfms}, matching AGN to non-AGN controls will not introduce a significant bias in stellar masses or SFRs that would lead to a misrepresentation of low-$z$ AGN.
    
    Each AGN is matched to as many non-AGN controls as fall within a 0.1 dex tolerance in M$_{\star\text{, total}}$, 0.1 dex tolerance in SFR$_\text{total}$, and 20\% of the physical radius of the 3" SDSS fibre. We require that at least 3 controls are matched, that at least 50\% of the matched controls have inner molecular gas detections with S/N $>10$, and that any inner non-detections present have an upper limit lower than the median of the controls. These requirements ensure that the median can be measured accurately while including upper limit values in the distribution. In cases where any of these three requirements are not met, we expand the control tolerances by 0.1 dex in M$_{\star\text{, total}}$, 0.1 dex in SFR and 10\% in physical radius until these requirements are met. We allow for a maximum of two grows, equivalent to tolerances of 0.3 dex in SFR and M$_{\star, \text{ total}}$ and 40\% of the inner radius. As a result of the inner radius match, controls are typically matched in redshift with a tolerance of $|\Delta z|<0.01$. Of the 81 AGN in SALVAGE, 70 of them are successfully matched to controls. We tested variations in the control matching methodology including changing the minimum required controls between two and five, varying the number of allowed grows from one to four, and requiring central gas detections in all non-AGN, and found that the qualitative results do not change.

    \subsubsection{Metrics for Comparing AGN and Controls}

    We use a suite of three metrics to compare the molecular gas in the central regions of AGN host to non-AGN control galaxies. First, for each AGN, we compute the difference between the molecular gas surface density of the inner region from non-AGN controls, $\Delta\Sigma_\text{mol, inner}$:

    \begin{equation}
        \Delta\Sigma_\text{mol, inner} = \log\left(\Sigma_\text{mol, inner, AGN}\right) - \log\left(\text{median}[\Sigma_\text{mol, inner, ctls}]\right),
    \end{equation}

    \noindent where $\Sigma_\text{mol, inner, AGN}$ is the molecular gas surface density in units of M$_\odot$ kpc$^{-2}$ of an AGN and $\text{median}[\Sigma_\text{mol, inner, ctls}]$ is the median molecular gas surface density of the matched controls.

    We compute a second metric to compare the molecular gas reservoir of the AGN to their controls, relative to the stellar mass in the inner region, $\Delta f_\text{gas, inner}$:

    \begin{equation}
        \Delta f_\text{gas, inner} = \log\left(f_\text{gas, inner, AGN}\right) - \log\left(\text{median}[f_\text{gas, inner, ctls}]\right)
    \end{equation}

    \noindent where $f_\text{gas, inner, AGN}$ is the gas fraction (unitless) of an AGN and $\text{median}[f_\text{gas, inner, ctls}]$ is the median gas fraction of the matched controls.

    Lastly, \citet{Ellison21} and \citet{Garcia-Burillo24} both found signals of central gas depletion in AGN using self-referential metrics. We therefore also test for central gas depletion using using $\nabla f_\text{gas}$, and how that varies compared to a matched controls with $\Delta \nabla f_\text{gas}$:

    \begin{equation}
        \Delta \nabla f_\text{gas} = \log\left(\nabla f_\text{gas, AGN}\right) - \log\left(\text{median}[\nabla f_\text{gas, ctls}]\right).
    \end{equation}

    We also compute $\Delta\Sigma_\text{mol, inner}$, $\Delta f_\text{gas, inner}$, and $\Delta \nabla f_\text{gas}$ for galaxies in the non-AGN control pool as an assessment of the typical galaxy-to-galaxy variation of these quantities for non-AGN.

%% file: S4_Results.tex
\section{Results}
\label{ch4-results}

\subsection{The CO Distributions of AGN Hosts}

    \begin{figure*}
        \centering
        \includegraphics[width=1\linewidth]{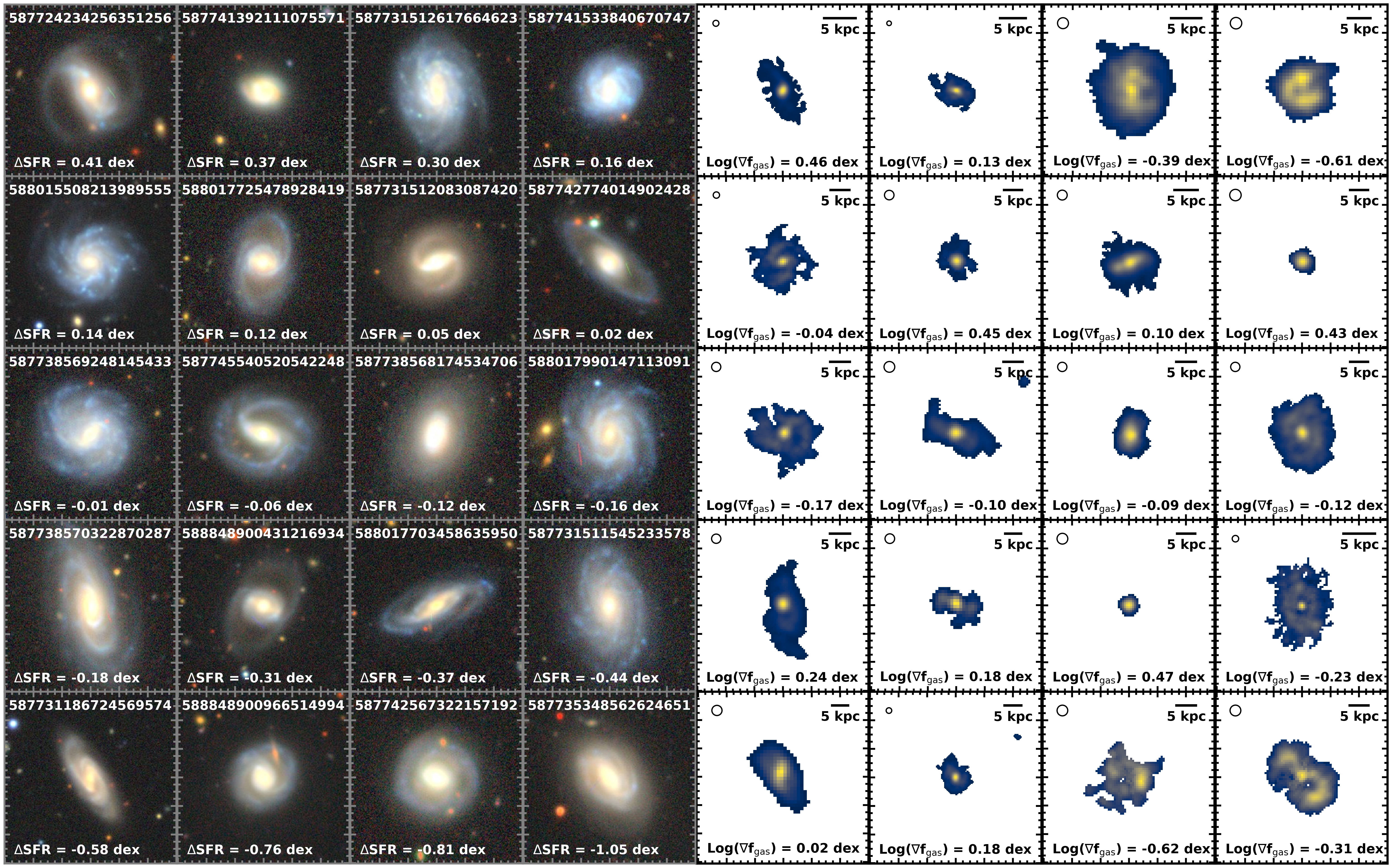}
        \caption[Images for 20 example AGN with strong CO detections]{Images for 20 example AGN with strong global CO detections. On the left are the 1' $\times$ 1' cutouts from the DESI-LS (centred on the SDSS fibre) and on the right are the CO moment 0 maps (centred on the ALMA observation RA and Dec. and with a scale matched to the 1 arcmin optical cutouts). The galaxies are ordered according to their $\Delta$SFR, which is listed in the bottom left of the optical images. At the top of the CO maps we show the beam size and a 5 kpc scale bar and along the bottom of the CO moment 0 maps, we report the value of $\log(\nabla f_\text{gas})$. The distribution of the full sample of $\log(\nabla f_\text{gas})$ is presented in Figure \ref{grad_fgas_agn_histo}.}
        \label{AGNexamples}
    \end{figure*}

    We begin our investigation of the AGN sample by broadly characterizing the AGN hosts and the distribution of their molecular gas. In Figure \ref{AGNexamples}, we present 20 examples of AGN with strong global CO detections. On the left side, we show 1 arcmin cutouts of the optical \emph{grz} colour images from the DESI-LS \citep{Dey19}. The optical images of the AGN show indications of ongoing interactions in some cases and undisturbed spiral arm structure in others. Many of the 20 example AGN also show indications of a strong stellar bar. On the right side of Figure \ref{AGNexamples}, we show the CO line emission maps of the AGN matched to the same angular scale as the optical images. In most cases, the AGN have a CO-bright centre. However, some AGN have smoothly distributed CO emission (e.g., centre panel) and even decreasing CO emission towards the centre (e.g., top right panel). Broadly speaking, the AGN host galaxies in SALVAGE tend to have a diversity of molecular gas distributions.

    \begin{figure*}
        \centering
        \includegraphics[width=\linewidth]{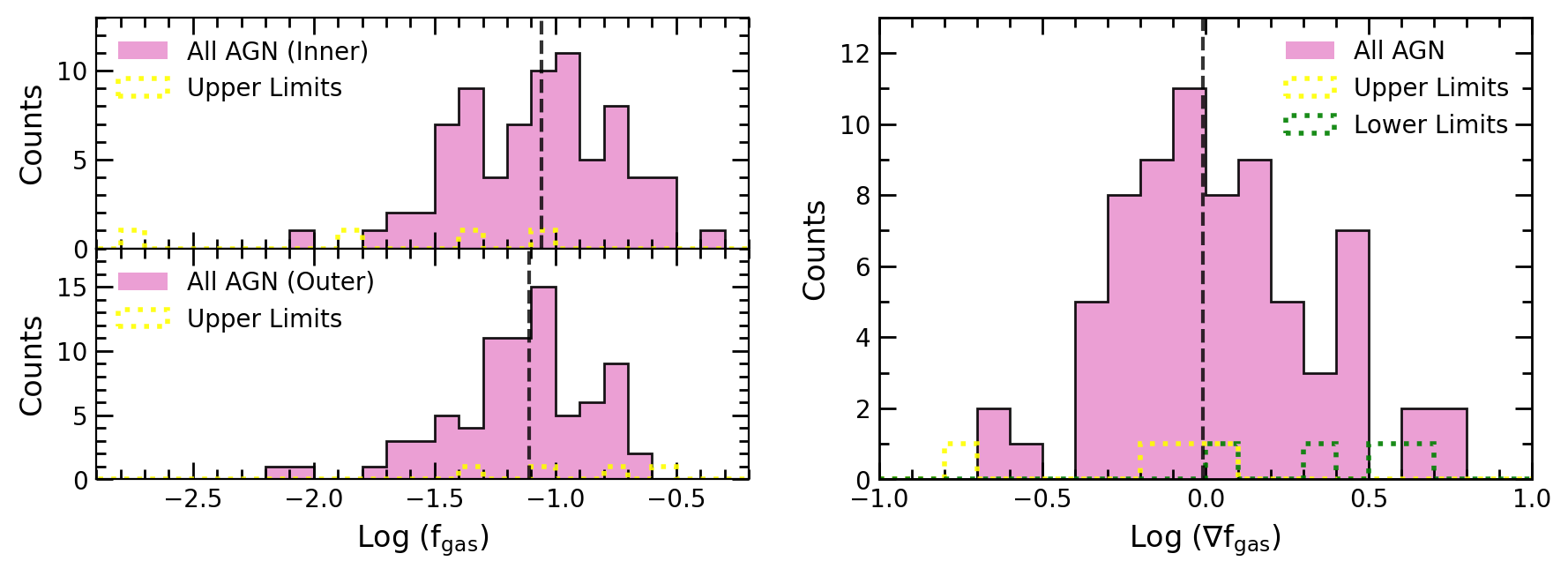}
        \caption{\emph{Left:} The distribution of log($f_\text{gas, inner}$) (top panel) and log($f_\text{gas, outer}$) (bottom panel) for all AGN in SALVAGE (including all AGN selection methods and before requiring successful control matching). The black dashed line represents the median of the distribution, excluding upper limits (yellow dotted distribution). \emph{Right:} The distribution of log($\nabla f_\text{gas}$), equivalent to log($f_\text{gas, inner} / f_\text{gas, outer}$), for all AGN in SALVAGE. The black dashed line shows the median of the AGN population (log($\nabla f_\text{gas}$) $=0$) excluding upper limits (yellow dotted distribution) and lower limits (green dotted distribution). Approximately equal numbers of AGN have higher/lower gas fractions in the central regions compared to their outer regions, and in general, there is a large diversity of molecular gas profiles.}
        \label{grad_fgas_agn_histo}
    \end{figure*}

    We parameterize the molecular gas profile of the AGN relative to the distribution of the stellar mass using the ratio of the inner and outer gas fractions, $\nabla f_\text{gas}$. In the left panel of Figure \ref{grad_fgas_agn_histo}, we show the distributions of log($f_\text{gas, inner}$) and log($f_\text{gas, outer}$) for all 81 AGN identified in SALVAGE. The medians of the distributions are shown as black dashed lines, which exclude galaxies for which only upper limits could be measured (dotted yellow distribution). In the right panel of Figure \ref{grad_fgas_agn_histo}, we show the distribution of log($\nabla f_\text{gas}$) for all 81 AGN in SALVAGE. Since either the inner or outer region may have a non-detection (but not both, by selection) there are both upper limits (yellow dotted distribution) and lower limits (green dotted distribution) on log($\nabla f_\text{gas}$).
    
    The median value of log($\nabla f_\text{gas}$) for the AGN sample is $0.00\pm0.04$ dex, indicating that the AGN host tends towards a flat gas fraction profile. Indeed, the individual medians of the inner (log($f_\text{gas, inner}$)$=-1.05\pm0.05$ dex) and outer (log($f_\text{gas, inner}$)$=-1.11\pm0.04$ dex) gas fraction are equal within error. The ratio between the inner and outer regions has a large diversity, including both very centrally depleted cases (log($\nabla f_\text{gas}$) $<-0.5$ dex) and centrally concentrated cases (log($\nabla f_\text{gas}$) $>+0.5$ dex). The gas fraction ratio of the entire sample demonstrates that the molecular gas distributions of AGN hosts are overall not too dissimilar from the stellar mass distributions, although they show a large variety of enhanced, normal, and deficient central gas reservoirs.

\subsection{The Central Molecular Gas Reservoirs of AGN Hosts}

\begin{figure*}
        \centering
        \includegraphics[width=1\linewidth]{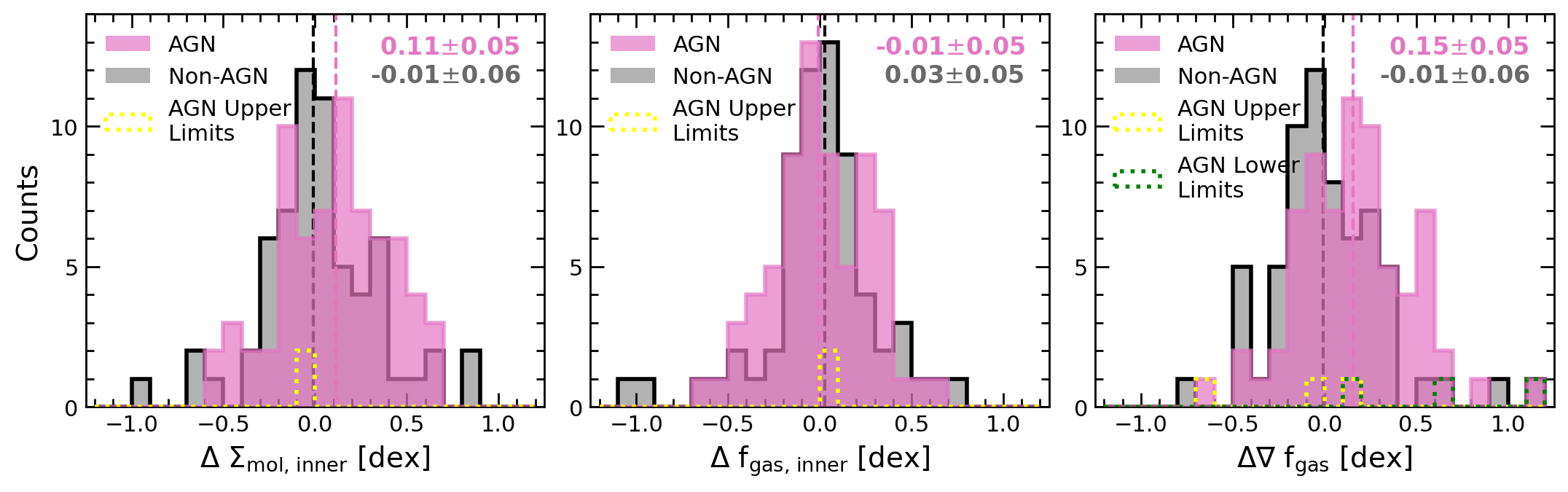}
        \caption[The distributions of $\Delta\Sigma_\text{mol, inner}$, $\Delta f_\text{gas, inner}$, and $\Delta \nabla f_\text{gas}$, each quantifying the difference between the central molecular gas reservoir of the AGN from the controls]{The distributions of $\Delta\Sigma_\text{mol, inner}$ (left panel), $\Delta f_\text{gas, inner}$ (middle panel), and $\Delta \nabla f_\text{gas}$ (right panel), each quantifying the difference between the central molecular gas reservoir of the AGN from the controls, are shown as pink histograms. Grey histograms represent the same metrics for non-AGN matched to their own non-AGN controls to demonstrate the typical variance of this metric for non-AGN. The median of the AGN distributions are shown as pink dashed lines and the non-AGN distributions are black, with the values of these median reported in the top right corner. Upper limits and lower limits are included in the histograms, but also shown separately as yellow and green dotted lines, respectively.}
        \label{deltas_all_agn}
    \end{figure*}

    Having established that the AGN sample has a diversity of molecular gas distributions, including central gas enhancements and central gas deficits, we now compare the central molecular gas reservoirs of the AGN hosts directly to their matched controls. In Figure \ref{deltas_all_agn}, we present the distribution of $\Delta\Sigma_\text{mol, inner}$, $\Delta f_\text{gas, inner}$, and $\Delta \nabla f_\text{gas}$ of the AGN as pink histograms. We also plot the distribution of these metrics for non-AGN matched to other non-AGN controls (grey histograms) to demonstrate the typical variance of these metrics, even for non-AGN. In the top right corner of each panel, we report the median of the AGN distribution in pink and the non-AGN in grey. By definition, the median of the non-AGN controls should always be consistent with 0. 

    Looking first at the left panel of Figure \ref{deltas_all_agn}, we find that the AGN distribution of $\Delta\Sigma_\text{mol, inner}$ is broad, again demonstrating the diversity of the AGN molecular gas properties. Using the median value of the distribution,  $\Delta\Sigma_\text{mol, inner} = (0.11\pm0.05)$ dex, we find a tentative enhancement of the central molecular gas mass surface densities of AGN host galaxies with a $2.2\sigma$ significance. The upper limits introduced by non-detections are less than the median and therefore their inclusion allows for a robust assessment of the median. As expected by the control matching scheme, the non-AGN population has a median consistent with 0 ($\Delta\Sigma_\text{mol, inner} = (-0.01\pm0.05$) dex). However, non-AGN also exhibit a broad range of extreme $\Delta\Sigma_\text{mol, inner}$ offsets from other non-AGN, as high as $\thicksim$1 dex above and below the median. The broad values of the non-AGN distribution demonstrates that the $\pm$0.5 dex change in $\Sigma_\text{mol, inner}$ seen in some AGN hosts is not unique to the AGN host population.

    In the centre panel of Figure \ref{deltas_all_agn}, we test if the AGN population has higher gas fractions in their inner regions compared to non-AGN controls. We find that the distribution of $\Delta f_\text{gas, inner}$ for the AGN is centred around 0 with a median of the AGN distribution is ($-0.01\pm0.05$) dex. The upper limits of the AGN are above the median. However, artificially moving them below the median (which could be true, based on the upper limits) does not change the median beyond its error and does not affect our conclusion: when normalizing by stellar mass in the inner region, the molecular gas enhancement seen in the molecular gas surface density goes away.

    Lastly, in the right panel of Figure \ref{deltas_all_agn}, we test if using the self-referential metric $\nabla f_\text{gas}$ shows a difference between the gas distribution in AGN compared to their controls. The median of the distribution of $\Delta \nabla f_\text{gas}$ is found to be ($0.15\pm0.05$) dex, which demonstrates a statistically significant enhancement (i.e., $3\sigma$) of the gas fraction gradient in AGN hosts relative to non-AGN controls. Since there are equal numbers of upper and lower limits above and below the median of the AGN distribution, the median would not change if we excluded them.
    
    \subsection{Comparing Different AGN Selection Methods}
    \label{agnresults-split}

    In the previous subsection, we found that, when considering the entire population of 70 AGN in SALVAGE, AGN have enhancements in $\Sigma_\text{mol, inner}$ and $\nabla f_\text{gas}$, but no significant enhancement (or depletion) in $f_\text{gas, inner}$. In this subsection, we now test if that result holds for all subpopulations of AGN.
    
\begin{figure}
        \centering
        \includegraphics[width=\linewidth]{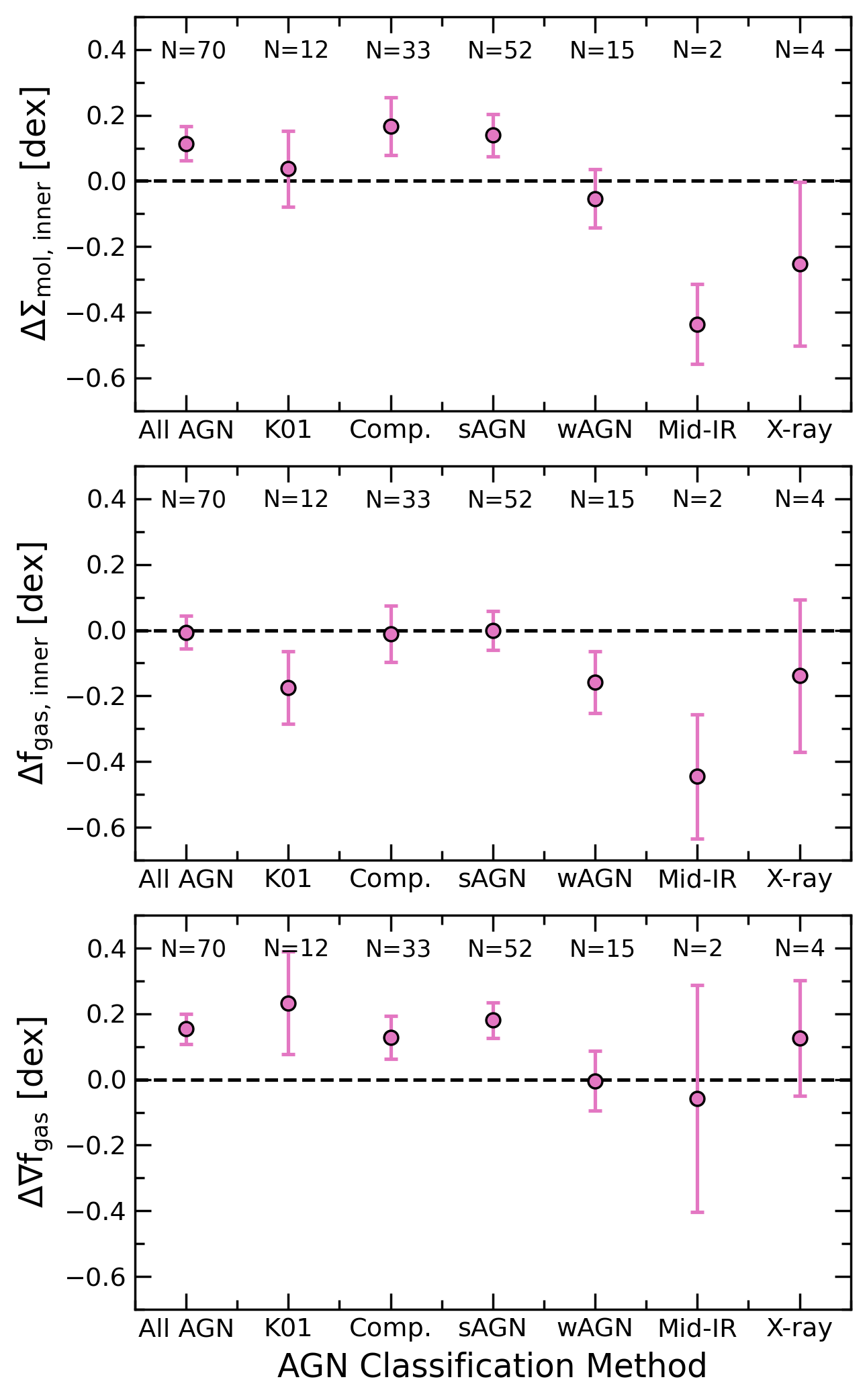}
        \caption[The median difference between AGN and their controls in $\Sigma_\text{mol, inner}$, $f_\text{gas, inner}$, and $\nabla f_\text{gas} $, split across the different AGN classification methods]{The median difference between AGN and their controls in $\Sigma_\text{mol, inner}$ (top panel), $f_\text{gas, inner}$ (middle panel), and $\nabla f_\text{gas} $ (bottom panel), split across the different AGN classification methods. The errorbars represent the standard error on the median and the black dashed line is set at zero to guide the eye. Above each point is the number of AGN classified by the method and successfully matched to controls. Regardless of the AGN classification method and measure of central gas, AGN typically have normal or enhanced central molecular gas compared to their controls.}
        \label{delta_fgas_agn_split}
    \end{figure}

    In Figure \ref{delta_fgas_agn_split}, we present the median $\Delta\Sigma_\text{mol, inner}$ (top panel), $\Delta f_\text{gas, inner}$ (centre panel), and $\Delta \nabla f_\text{gas}$ for each of the AGN classification methods. The errors on the points represent the standard error on the median and the black dashed line at zero represents no change from the non-AGN control population. Along the top of each panel are the number of AGN included in each sample; the sum of AGN subsets is larger than the ``all AGN'' sample because AGN can be identified by multiple methods. For reference, the ``All AGN'' medians are the exact same as reported in Figure \ref{deltas_all_agn}.

    The top panel of Figure \ref{delta_fgas_agn_split} reveals that the median enhancement of $\Sigma_\text{mol, inner}$ in the SALVAGE AGN sample is largely driven by the composite AGN and sAGN subpopulations. Indeed, sAGN make up 74\% of the total AGN sample and have a median enhancement of $\Delta\Sigma_\text{mol, inner} = (0.14 \pm 0.06)$ dex. In contrast, the K01 AGN and wAGN samples are smaller ($N=12$ and $N=15$, respectively), but have no statistical enhancement in $\Sigma_\text{mol, inner}$. The mid-IR AGN have a median offset of $\Delta\Sigma_\text{mol, inner} = (-0.44 \pm 0.12)$ dex, but since the sample size is small ($N=2$), we refrain from over-interpreting the significance of this result. 

    In the middle panel of Figure \ref{delta_fgas_agn_split}, we test how $\Delta f_\text{gas, inner}$ varies across the different AGN populations. We find that the lack of enhancement/depletion in the whole AGN population is also driven by the composite AGN and sAGN ($N=33$ and $N=52$, respectively); both of these subpopulations have median $\Delta f_\text{gas, inner}$ consistent with 0. However, looking at the K01 AGN and wAGN subpopulations, we find a $1.5\sigma$ and $1.8\sigma$ signal, respectively, hinting towards centrally depleted gas fractions. The K01 AGN population has a median of $\Delta f_\text{gas, inner} = (-0.17 \pm 0.11)$ dex and the wAGN has a median of $\Delta f_\text{gas, inner} = (-0.16 \pm 0.09)$ dex.

    In the bottom panel of Figure \ref{delta_fgas_agn_split}, we present the median values of $\Delta \nabla f_\text{gas}$ for each of the AGN classification methods. In this case, we find composite AGN, sAGN, as well as K01 AGN have significantly enhanced $\nabla f_\text{gas}$, and are broadly representative of the AGN sample as a whole. However, the wAGN, mid-IR AGN, and X-ray AGN have median values consistent with no change from the control sample. In this test, we find that the difference in the K01 and wAGN populations do lead to different conclusions where they did not in previous tests.

    By splitting the AGN into different subpopulations, we have revealed that the results in the previous subsection are driven by the more numerous composite AGN and sAGN samples. In contrast, the K01 AGN and wAGN samples behave similar to each other, but different from the composite and sAGN samples. Namely, the median $\Sigma_\text{mol, inner}$ of the K01 AGN and wAGN are equal (not enhanced) relative to non-AGN controls and the median values of $f_\text{gas, inner}$ are depleted (not normal) relative to non-AGN controls. The mid-IR and X-ray AGN samples are too small to make conclusions representative of their AGN subpopulations.

    \subsection{Searching for Trends with Galaxy Properties}

    In the previous subsection, we found 1.5$\sigma$ and $1.8\sigma$ signals that hint towards a possible $f_\text{gas, inner}$ depletion in the K01 AGN and wAGN populations compared to non-AGN controls. To explore this further, we will investigate if $f_\text{gas, inner}$ trends with key galaxy properties. This test allows us to better understand if there is a particular regime in which galaxies may be more susceptible to central gas depletion. Moreover, if K01 AGN and wAGN occupy similar regimes of certain galaxy properties, this could reveal the underlying physics at play in these AGN. 

\begin{figure*}
        \centering
        \includegraphics[width=\linewidth]{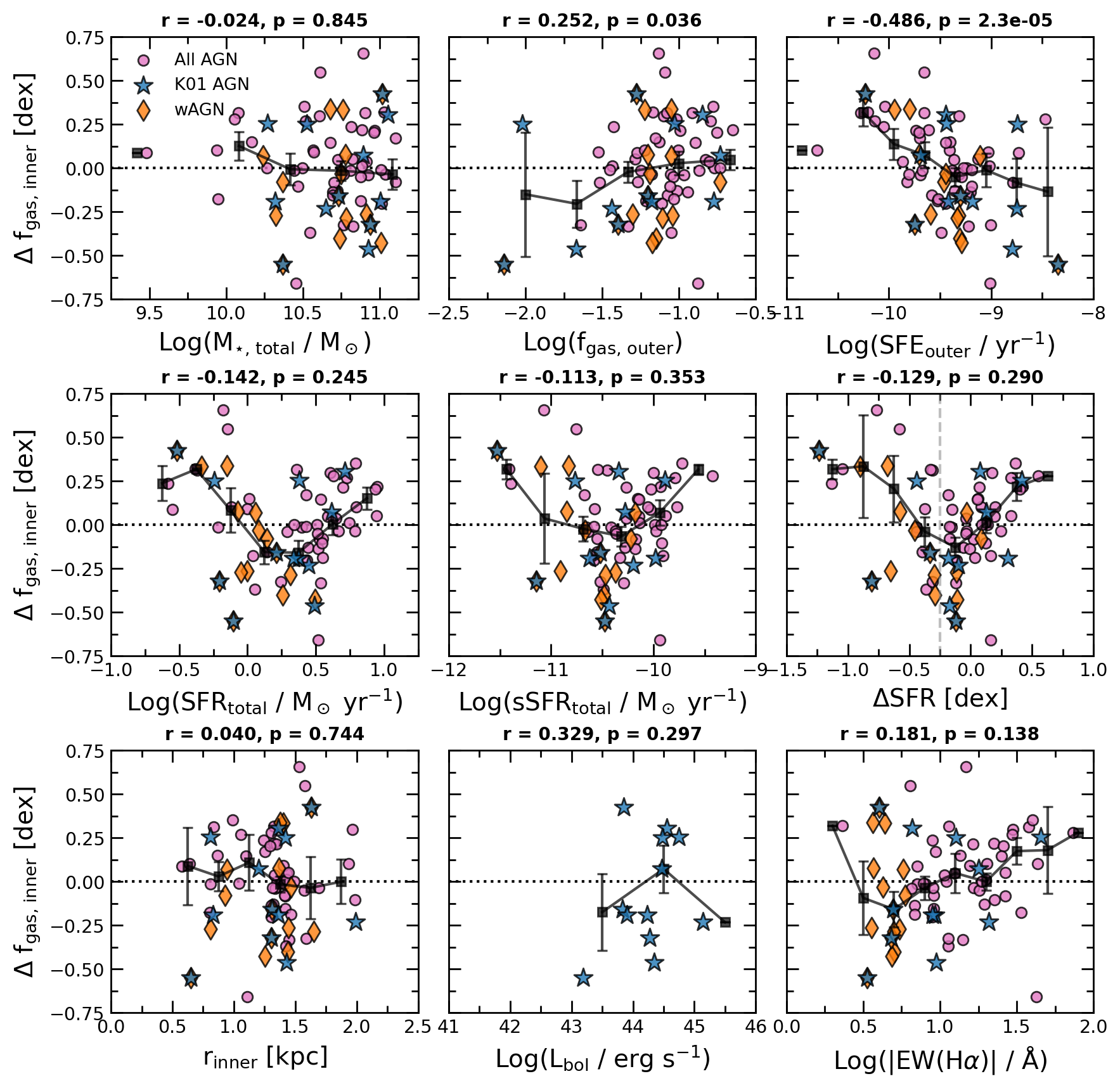}
        \caption{The difference between the central gas fraction of AGN and their controls, $\Delta f_\text{gas, inner}$, as a function of various host galaxies properties. In each panel, the AGN sample is shown as pink circles, except for K01 AGN (blue stars) and wAGN (orange triangles) to identify any reason these AGN classifications have a negative median $\Delta f_\text{gas, inner}$. To guide the eye, there is a black dotted line at $\Delta f_\text{gas, inner}$ (representing no difference from matched controls) and the running median of the AGN sample is shown as black squares. The Pearson correlation coefficient ($r$) and its associated p-value is given at the top of each panel.}
        \label{agn_trends}
    \end{figure*}

    In Figure \ref{agn_trends}, we present the trend between $\Delta f_\text{gas, inner}$ and nine different galaxy properties. In each panel, we plot $\Delta f_\text{gas, inner}$ against the galaxy property for the entire AGN sample (pink circles). We highlight K01 AGN (blue stars) and wAGN (orange diamonds) in particular to test if they occupy a unique and similar part of the parameter space. To guide the eye, we plot a running median (black squares) with the error on the median represented by the black error bars. At the top of each panel, we report the Pearson correlation coefficient and its p-value as a measure of the linear correlation between the variables. For this test, we have removed one significant outlier (SDSS ObjID 587738568174731329) as its extreme value of $\Delta f_\text{gas, inner}$ ($-1.73$) strongly dictated the correlation coefficients. The properties in each row of panels are loosely organized into themes: global galaxy properties, measurements of SFR, and AGN properties.

    \subsubsection{Global Galaxy Properties}

    We begin by investigating the global properties of the AGN host galaxies whose correlations with $\Delta f_\text{gas, inner}$ are presented in the top row of Figure \ref{agn_trends}. For global properties that include M$_\text{mol, inner}$ in their computation (i.e. $f_\text{gas, total}$ and SFE$_\text{total}$, we focus on only the independent outer regions to avoid direct correlations with $\Delta f_\text{gas, inner}$.
    
    For properties that include the molecular gas  In the left panel of the top row in Figure \ref{agn_trends}, we present the trend between $\Delta f_\text{gas, inner}$ and M$_{\star\text{, total}}$. Based on the flat median trend and significant scatter around $\Delta f_\text{gas, inner} = 0$, we find no significant trend with stellar mass. Indeed the Pearson correlation coefficient ($r = -0.024, p = 0.845$) reveals no significant correlation. Thus, the total stellar mass has no bearing on the inner central gas depletion of AGN in SALVAGE.

    In the top centre panel, we show the trend between $\Delta f_\text{gas, inner}$ and the $f_\text{gas, outer}$ of the AGN host galaxy. We use $f_\text{gas, outer}$ as a measurement of the global gas reservoir that is still measured independently of $\Delta f_\text{gas, inner}$ (as opposed to using $f_\text{gas, total}$ which is not independent from $\Delta f_\text{gas, inner}$). We find a large scatter, but a weak positive trend ($r = 0.252, p = 0.036$). Therefore, the $\Delta f_\text{gas, inner}$ offset seems to have a minimal connection to the total gas reservoir available in the system. 

    In the top right panel, we show the trend between $\Delta f_\text{gas, inner}$ and the SFE$_\text{outer}$ of the host galaxy. We find a moderate negative trend ($r = -0.486$) that is statistically significant ($p = 2.3\times10^{-5}$). In particular, galaxies with low SFE tend to have enhanced $f_\text{gas, inner}$ relative to non-AGN controls. Conversely, AGN host galaxies with low $\Delta f_\text{gas, inner}$ are often found galaxies that are more efficient in their outer regions.

    In all three cases, the K01 AGN and wAGN do not occupy a unique part of the parameter space.

    \subsubsection{Measurements of SFR}

    In the centre row of Figure \ref{agn_trends}, we plot the trend between $\Delta f_\text{gas, inner}$ and three measurements of SFR: SFR$_\text{total}$, sSFR$_\text{total}$, and $\Delta$SFR. All three show a distinct ``U'' shape trend and no significant linear correlation (i.e. no trends with $p<0.05$). AGN hosts at the extremes of low and high SFR tend to have high $\Delta f_\text{gas, inner}$. In contrast, many of the AGN with low $\Delta f_\text{gas, inner}$ are clustered at intermediate and low SFRs at the bottom of the SFMS ($-0.5\text{ dex}< \Delta\text{SFR}<0.0\text{ dex}$).

    The ``U'' shape trend seen with $\Delta$SFR is statistically significant. If we compute the linear correlation coefficient twice, once for the high $\Delta$SFR regime ($\Delta \text{SFR} > -0.25$ dex) and once for the low $\Delta$SFR regime ($\Delta \text{SFR} < -0.25$ dex), the two linear correlations are moderate to strong ($r = +0.512$ and $r = -0.489$, respectively) and significant ($p = 2.7\times10^{-4}$ and $p = 0.018$, respectively). The threshold of $\Delta \text{SFR} = -0.25$ dex is arbitrary (see dashed grey line in Figure \ref{agn_trends}) and significant correlation coefficients are found using any threshold in the bottom half of the SFMS $-0.3 \text{ dex}< \Delta \text{SFR} < 0.0\text{ dex}$.

    Since the AGN host galaxies are matched in M$_{\star\text{, total}}$ and SFR$_\text{total}$, the positive correlation between $\Delta f_\text{gas, inner}$ and $\Delta \text{SFR}$ is not simply a recreation of the KS relation. Instead, the change in central gas fraction coincident with the AGN dictates the position of the host galaxy on the SFMS. This trend is consistent with a physical scenario in which a central gas enhancement that triggers an AGN coincides with a burst of star formation and separately or subsequently a decrease in inner gas fraction leading to suppressed star formation. However, since the positive correlation stops around $\Delta \text{SFR} = -0.25$ dex, there is no evidence that gas removal by AGN leads to quenching (as defined by leaving the SFMS, i.e. $\Delta \text{SFR} < -0.3$ dex). The turnover and negative correlation below the SFMS may be caused by a radical change in the control sample, where most galaxies become gas poor.

    In all three cases, the K01 AGN and wAGN do not occupy a unique part of the parameter space.

    \subsubsection{AGN-related Properties}

    In the bottom row of Figure \ref{agn_trends}, we plot the trend between $\Delta f_\text{gas, inner}$ and three metrics relevant to AGN feedback in the inner region. In the left panel, we test if our assessment of $\Delta f_\text{gas, inner}$ is sensitive to the physical size of the inner region measurement, as defined by the 3" SDSS fibre. We find no significant trend between $\Delta f_\text{gas, inner}$ and $r_\text{inner}$ indicating that our conclusion are consistent across the physical scales probed by the SALVAGE data ($0.5 \text{ kpc}<r_\text{inner}<2 \text{ kpc}$. Moreover, K01 AGN and wAGN do not occupy a unique part of the parameter space.
    
    In the bottom centre panel, we test if there is a correlation between the central molecular gas and AGN luminosity. The bolometric luminosity of the AGN ($L_\text{bol}$) is measured using the luminosity of the dust-corrected [OIII] emission line and the bolometric conversion derived for SDSS galaxies in \citet{Heckman04}, which has a typical uncertainty of 0.38 dex for a given AGN. We conduct this test only on K01 AGN where line emission contributions from star forming regions are minimal. We find no significant correlation between $\Delta f_\text{gas, inner}$ and $L_\text{bol}$.

    Since the only difference between wAGN (which have $\Delta f_\text{gas, inner} < 0$) and sAGN (which have $\Delta f_\text{gas, inner} \simeq 0$) is the EW(H$\alpha$), we test if EW(H$\alpha$) correlates with $\Delta f_\text{gas, inner}$ in the bottom right panel. Although the median values of $\Delta f_\text{gas, inner}$ increase with increasing EW(H$\alpha$), there is no statistically significant correlation between $\Delta f_\text{gas, inner}$ and EW(H$\alpha$). By definition, wAGN are found at log(EW(H$\alpha$))$<0.78$, but K01 AGN span the entire parameter space.

%% file: S5_Discussion.tex
\section{Discussion}
\label{ch4-discussion}

In this work, we have used SALVAGE molecular gas observations from the ALMA archive to explore a statistically large sample ($N=70$) of diverse AGN (six different selection criteria) at a spatial resolution of $r\lesssim $ 0.5-2 kpc to test the impact of AGN feedback on the central molecular gas reservoir. By virtue of the diverse assortment of AGN in SALVAGE, our results average over many important factors including the strength of the AGN (and its ability to couple to the ISM), the timescale of AGN observability compared to gas removal, and radial extent over which AGN may impact the ISM. Indeed, consolidating the 70 AGN tested in this work, we have found no signal of systematic molecular gas depletion in the central regions of AGN (see Figure \ref{delta_fgas_agn_split}). Must we conclude that AGN do not have an impact on the molecular gas reservoirs, or can we reconcile our results with the impact of these factors?%In this section, we will discuss the entanglement of these individual factors and how they relate to previous works in the literature. 

Studies of small samples of AGN have led to mixed conclusions when testing molecular gas depletion at kpc scales. For example, using kpc-scale EDGE-CALIFA observations of K01 AGN has led to both signals of AGN depletion and no difference from controls, depending on the methodology \citep{Ellison21, Yu22}. In another work, \citet{Rosario18} study 18 X-ray-selected AGN ($L_X=10^{42-43}$ erg s$^{-1}$) and demonstrate that there is no statistical difference in the central kpc molecular gas content of the AGN as compared to an inactive control sample matched in H-band luminosity (similar to our mass match), Hubble type (similar to our SFR match), and inclination. \textbf{At the same X-ray luminosities ($L_{X }> $ 42 erg s$^{-1}$), the central molecular gas of AGN hosts has been shown to be depleted at 50 pc scales} \citep{Maccagni18, Audibert19, AA18, AA23, Garcia-Burillo24}. At similarly high AGN luminosities, \citet{RA22} studied 7 QSOs ($L_\text{bol}>10^{45.5}$ erg s$^{-1}$) with a resolution of 0.2" (370 pc) and found a mix of  centrally-peaked and centrally-depleted molecular gas CO profiles. \citet{Elford24} and \citet{Garcia-Burillo24} also test central molecular gas depletion at 50 pc but for less luminous AGN and find no difference to non-AGN. Therefore, clear signals of AGN depletion have only been found at high luminosities and at 50 pc scales where AGN observability and gas depletion timescales are comparable. 

The SALVAGE AGN sample fills in the 0.5 to 2 kpc regime between \citet{RA22} and \citet{Rosario18}, with preferentially lower AGN luminosities (only one X-ray AGN above $L_X>10^{42}$ erg s$^{-1}$ and one K01 AGN above $L_\text{bol}>10^{45}$ erg s$^{-1}$) but larger sample statistics ($N=70$). At all resolutions probed in this work (0.5 to 2 kpc), we find that the molecular gas fractions in the central regions are similar to their matched non-AGN controls and no significant correlation with AGN luminosity (see bottom row of Figure \ref{agn_trends}). \textbf{Our work has therefore found that averaging over a large number of diverse AGN supports the conclusions of \citet{Rosario18}}. In Section \ref{agnresults-split}, we found that K01 AGN and wAGN show signals of central gas depletion when using the $\Delta f_\text{gas}$ metric. However, investigating the trends of $\Delta f_\text{gas}$ with different galaxy properties with K01 AGN and wAGN in mind, revealed no particular regime in which galaxies may be more likely to have central gas depletion. This indicates that the reason for their depletion is not captured in the properties to which we have access. 

One parameter that we do not have access to is a timescale of the AGN event. AGN are thought to ``flicker'' between periods of high and low accretion on the order of $\thicksim 10^5$ yr \citep{Hickox14, Schawinski15}. Many studies have concluded that the on/off cycle of AGN feedback may be shorter lived than the observable impact on the gas reservoir, leading to the idea of ``fossil'' outflows \citep{Fluetsch19, Zubovas23, Alban24} where the outflow from an AGN is observed, but the AGN is no longer active. The disconnection between the timescales over which AGN emission may be observed, AGN accretion/feedback is occurring, and the observable effect on the ISM is likely contributing noise to studies observing AGN impact on the ISM. For one, we may be observing AGN that are actually off, as was observed in a case study by \citet{Ichikawa19}. On the other hand, a non-AGN galaxy could be still be experiencing the impact of AGN feedback, even though the AGN is no longer active, as was observed in a case study by \citet{Veronese25}. \textbf{If AGN events do indeed lead to wide-spread gas depletion, our work has contributed an upper limit ($r\lesssim 1$ kpc) on the scale over which that depletion can take place during the timescale of AGN observability.}

Although various types of AGN have global molecular gas experiments, including QSOs \citep[e.g.,][]{Scoville03, Bertram07, Shangguan-CO21-QSO-survey}, radio AGN \citep[e.g.,][]{OcanaFlaquer10, Tadhunter24}, X-ray AGN \citep[e.g.,][]{Rosario18, Koss21}, and optically-selected AGN \citep[e.g.,][]{Saintonge17, Bazzi25}, these studies are typically dedicated single dish observations of tens of objects. No one has previously tested the \emph{resolved} molecular gas of multiple AGN types in a homogeneous assessment. The work presented here is the first to internally compare multiple AGN selection methods, including mid-IR AGN which have never been studied in CO(1-0). We find that composite AGN (from the BPT diagram) and sAGN (from the WHaN diagram) have enhanced molecular gas surface densities and normal gas fractions in their central regions, when compared to non-AGN controls. In contrast, K01 AGN (from the BPT diagram), wAGN (from the WHaN diagram), and mid-IR AGN (although only 2 examples with controls) have depleted molecular gas fractions in the central 2 kpc regions. Since the K01 AGN and wAGN are not particularly luminous (see Figure \ref{agn_trends}), it may be that these AGN criteria capture AGN at different times in the fueling/feedback cycle, as has been suggested in some previous works \citep[e.g.,][]{Schawinski07, Leslie16, Le25}. 

Throughout this work, we have assumed a constant CO-to-$\text{H}_2$ conversion factor, $\alpha_\text{CO}$, for the inner and outer regions of AGN and non-AGN alike. Previous works have shown that both AGN and non-AGN may have lower $\alpha_\text{CO}$ values ($\alpha_\text{CO} \thicksim 2$ (K km s$^{-1}$ pc$^{2}$)$^{-1}$) in their centres \citep{Sandstrom13, Garcia-Burillo14, Fei23} and variable $\alpha_\text{CO}$ (possibly higher or lower than 4.35 M$_\odot$ (K km s$^{-1}$ pc$^{2}$)$^{-1}$) in the outer disk \citep{Esposito24}. 

Assuming a constant conversion factor may have several consequences in our analysis. For example, if $\alpha_\text{CO}$ is lower in the centre of AGN hosts than in the centre of non-AGN (say, due to increased nuclear turbulence), it would artificially increase the central molecular gas mass estimate and either \textit{weaken} any signal of molecular gas depletion or cause a misleading central gas enhancement. Specifically, to produce the 0.11 dex enhancement of $\Sigma_\text{mol, inner}$ compared to controls, the AGN $\alpha_\text{CO}$ in the inner regions would need only to be $\thicksim$3.4 M$_\odot$ (K km s$^{-1}$ pc$^{2}$)$^{-1}$ (i.e. using a 0.11 dex or 29\% overestimate of $\alpha_\text{CO}$ in AGN could explain our results).
 
We tested our results using the metallicity-dependent and stellar mass surface density-dependent $\alpha_\text{CO}$ prescription from \citet{Schinnerer24}. We found that the variable $\alpha_\text{CO}$ prescription lowered the value of $\alpha_\text{CO}$ in the inner regions of AGN and non-AGN to median values of 0.76 (K km s$^{-1}$ pc$^{2}$)$^{-1}$ and 0.83 (K km s$^{-1}$ pc$^{2}$)$^{-1}$, respectively (for more details see Appendix \ref{co-to-h2}). Accordingly, the value of $\Sigma_\text{mol, inner}$ decreased from $0.11\pm0.05$ dex to $0.07\pm0.05$ and thus the relative comparison between AGN and non-AGN was equal within error. Moving forward, resolved maps of multiple CO lines including CO isotopologues and several excitations \citep[e.g.][]{Teng23} for a large sample of AGN and non-AGN controls is critical to understanding if AGN systematically deplete central molecular gas reservoirs.

%% file: S6_Summary.tex
\section{Summary}
\label{ch4-summary}

   AGN feedback has been suggested as a dominant galaxy quenching mechanism through gas heating or ejection \citep{Man18, Harrison24}. However, the overwhelming majority of low-redshift studies of the global molecular gas reservoirs of AGN find that they are similar or gas-enhanced relative to inactive controls \citep[e.g.,][]{Shangguan20, Koss21, RA22, Salvestrini22, Molina23, Salome23}. There is nevertheless disagreement on the impact of AGN feedback on the molecular gas reservoirs at 0.5-2 kpc scales \citep{Rosario18, Molina21, Ellison21, Yu22, RA22}, although these previous works have been limited to small sample sizes ($N<20$).
   
   In this work, we use the SALVAGE sample, which provides independent measurements of the star-formation rate, stellar mass, and molecular gas mass of the inner 0.5-2 kpc, and the outer region of the galaxy. From SALVAGE, we select 70 AGN at $z<0.07$ using a suite of multi-wavelength AGN diagnostics: 33 composite AGN and 12 K01 AGN selected from the optical BPT diagram \citep{Baldwin81}; 52 sAGN and 15 wAGN selected from the optical WHaN diagram \citep{CidFernandes11}; 2 mid-IR AGN selected with a colour from WISE photometry \citep{Stern12, Blecha18}, and 4 X-ray selected AGN from archival X-ray observations \citep{Agostino19}. We inspect the central molecular gas reservoirs of the AGN compared to SFR-, M$_\star$-, and $z$-matched controls and reach the following conclusions: 

    \begin{itemize}

        \item In general, \textbf{AGN hosts do not have depleted molecular gas reservoirs in the central 0.5-2 kpc.} Visual inspection of the molecular gas maps of AGN host galaxies reveals a diversity of gas morphologies, including both centrally depleted and centrally enhanced molecular gas (see Figs. \ref{AGNexamples} and \ref{grad_fgas_agn_histo}). When compared to non-AGN controls matched in stellar mass, redshift, and SFR, we find that AGN hosts have \emph{enhanced} central molecular gas surface density (median $\Delta\Sigma_\text{mol, inner} = 0.12\pm0.05$) but comparable central gas fractions (median $\Delta f_\text{gas, inner} = -0.02\pm0.04$; see Fig. \ref{deltas_all_agn}). 
        \smallskip

        \item \textbf{K01 AGN and WHaN wAGN have lower $f_\text{gas, inner}$ than their non-AGN controls, but equal $\Sigma_\text{mol, inner}$}. By splitting the AGN population in SALVAGE into different selection criteria, we find that the molecular gas enhancements in the whole AGN population are driven by the more numerous composite AGN and sAGN. The K01 AGN and WHaN wAGN have lower gas fractions than matched controls (medians $\Delta f_\text{gas, inner} = (-0.17 \pm 0.11)$ dex and $\Delta f_\text{gas, inner} = (-0.16 \pm 0.09)$ dex, respectively), but equivalent molecular gas surface densities (medians $\Delta\Sigma_\text{mol, inner} = (0.04 \pm 0.12)$ dex and ($-0.05 \pm 0.09$) dex, respectively; see Fig. \ref{delta_fgas_agn_split}).
        \smallskip

        \item \textbf{There is no correlation between $\Delta f_\text{gas, inner}$ and several key parameters, such as AGN luminosity}. $\Delta f_\text{gas, inner}$ does not correlate with stellar mass ($r = -0.024, p = 0.845$), the size of the central aperture ($r = 0.04, p = 0.744$), or the luminosity of the AGN measured by the [OIII] line emission of K01 AGN ($r = 0.329, p = 0.297$). $\Delta f_\text{gas, inner}$ does correlate with $f_\text{gas, outer}$ and anti-correlate with SFE$_\text{outer}$, indicating that AGN with enhanced central gas fractions are often found in gas rich galaxies that are inefficiently forming stars in their outer regions (see Fig. \ref{agn_trends}).
        \smallskip

    \end{itemize}

    Taken together, we find no clear evidence that AGN systematically lead to molecular gas depletion at scales of 0.5-2 kpc. This may be a result of several observational limitations. For example, the radial extent of AGN depletion may be smaller than the resolution of SALVAGE data. Furthermore, our results are complicated by the timescale of large-scale depletion. Namely, wide-spread central gas depletion is unlikely while the AGN is “on” and after the AGN has removed gas, AGN-like emission may no longer be observable. To continue to develop our understanding of the effect AGN feedback has on the ISM, future studies should prioritize large and diverse sample sizes with higher-resolution observations of the central 500 pc, where AGN feedback on molecular gas is more likely to occur simultaneously with observable nuclear activity.

%% file: AppendixA.tex
\section{Stacking the SALVAGE ALMA Cubes}
\label{stackingcheck}

\begin{figure*}
        \centering
        \includegraphics[width=1\linewidth]{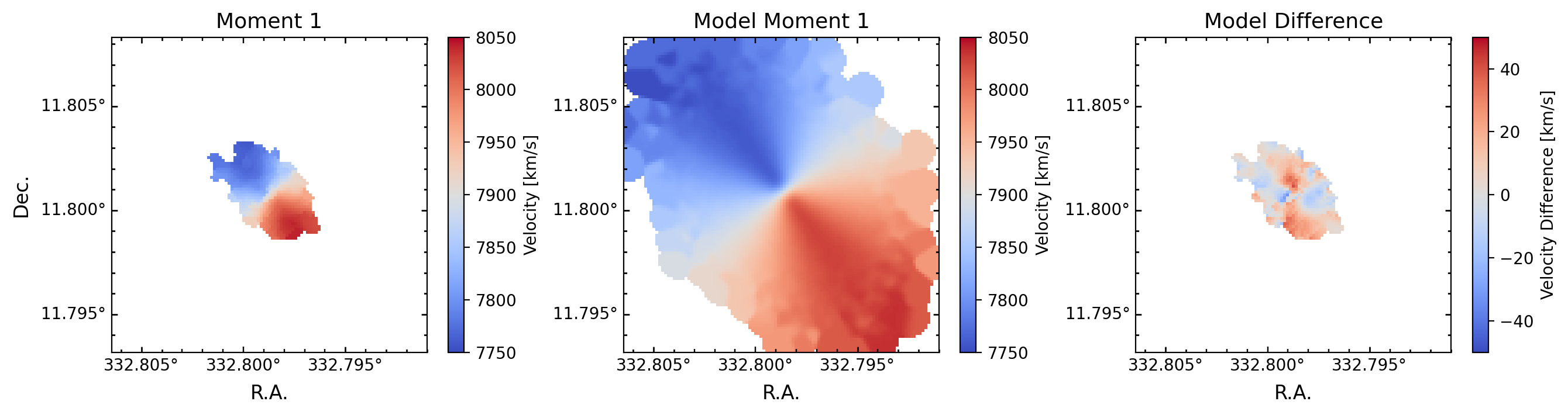}
        \caption[A comparison between the real and synthetic moment 1 map for SDSS Object 587727220865040712.]{A comparison between the real and synthetic moment 1 map for SDSS Object 587727220865040712. The left panel shows the moment 1 map from the PHANGS pipeline. The centre panel shows the moment 1 map created using \texttt{KinMS} and the optical morphology measurements from SDSS. The right panel shows the difference between these two maps. }
        \label{momentmapcompare}
    \end{figure*}

To further explore the reliability of the model moment 1 maps created using an exponential disk model and informed by only optical information, we directly compare one example of a model moment 1 map to a moment map measured from the ALMA cube in Figure \ref{momentmapcompare}. In the left panel, we show the moment 1 map output by the PHANGS pipeline for SDSS ObjID 587727220865040712. In the centre panel, we show the moment 1 map created using \texttt{KinMS}. In the right panel, we show the difference between the two. We find that for this galaxy, the maximum deviation between the model and the measured velocity profile is 40.4 km/s, or the equivalent of about 2 channels. In principle, as long as the maximum deviation between the model and the real CO velocity profile is less than the rotation velocity of the system (200-300 km/s), stacking with this method will improve S/N over not stacking at all \citep[as in][]{Wilkinson26}.

%% file: AppendixB.tex
\section{Testing a Variable $\alpha_\text{CO}$}
\label{co-to-h2}

In this section, we test how the results of our analysis would change if we were to use the variable $\alpha_\text{CO}$ prescription from \citet{Schinnerer24}. The \citet{Schinnerer24} prescription includes a metallicity-dependent factor $f(Z)$ and a stellar mass surface density-dependent factor $g(\Sigma_\star)$ to adjust the Milky Way-derived CO-to-H$_2$ conversion factor $\alpha_\text{CO, MW}$:

\begin{equation}
    \alpha_\text{CO} = \alpha_\text{CO, MW }f(Z)\text{ } g(\Sigma_\star)  
\end{equation}

\noindent where $f(Z)$ and $g(\Sigma_\star)$ are defined as follows:

\begin{equation}
    f(Z) = \left(\frac{Z}{Z_\odot}\right)^{-1.5}\text{, and}
\end{equation}

\begin{equation}
    g(\Sigma_\star) = \left(\frac{\text{max}(\Sigma_\star,100\text{ M}_\odot\text{ pc}^{-2})}{100\text{ M}_\odot\text{ pc}^{-2}}\right).
\end{equation}

Since the emission lines in the SDSS central fibres of AGN host galaxies are often not existent or contaminated by AGN emission, we cannot measure the gas phase metallicity of the AGN. Therefore, we estimate the metallicity from the mass-metallicity relation in \citet{Tremonti04}. We measure the stellar mass surface density directly from $M_{\star\text{, inner}}$ and the physical area contained within the SDSS fibre. For each galaxy in SALVAGE, we compute a variable $\alpha_\text{CO}$ for the inner region, and leave the outer region as $\alpha_\text{CO, MW} = 4.35$ M$_\odot$ (K km s$^{-1}$ pc$^{2}$)$^{-1}$.

\begin{figure*}
        \centering
        \includegraphics[width=1\linewidth]{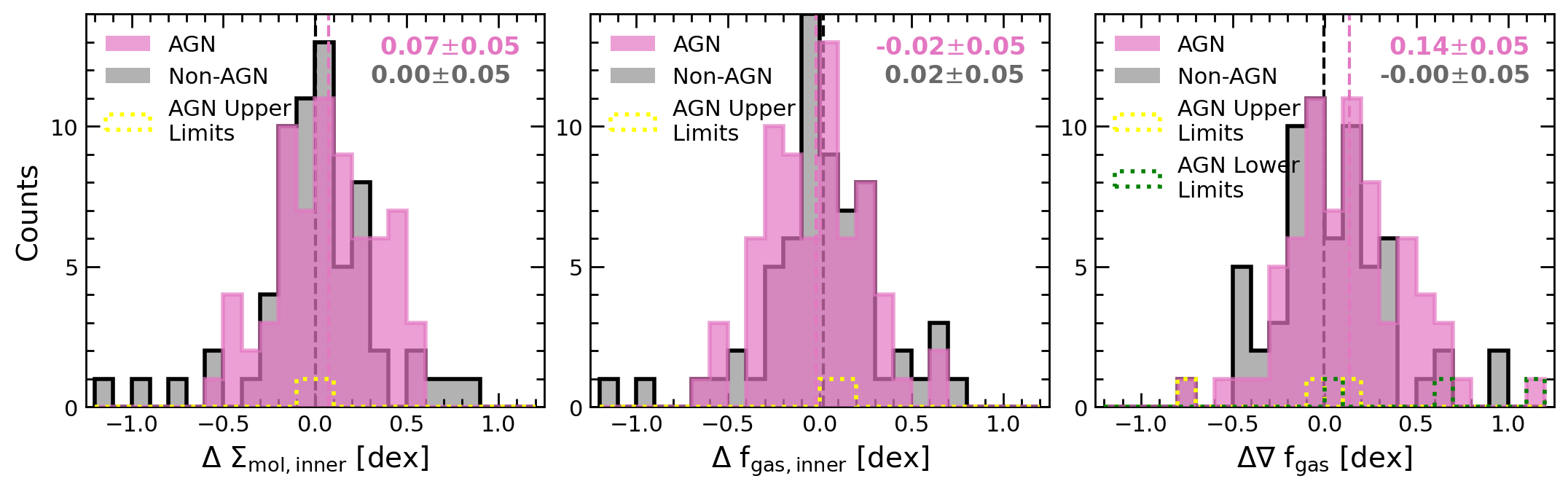}
        \caption{The distributions of $\Delta\Sigma_\text{mol, inner}$ (left panel), $\Delta f_\text{gas, inner}$ (middle panel), and $\Delta \nabla f_\text{gas}$ (right panel) when using the $\alpha_\text{CO}$ prescription from \citet{Schinnerer24}. Grey histograms represent the same metrics for non-AGN matched to their own non-AGN controls to demonstrate the typical variance of this metric for non-AGN. The median of the AGN distributions are shown as pink dashed lines and the non-AGN distributions are black, with the values of these median reported in the top right corner. Upper limits and lower limits are included in the histograms, but also shown separately as yellow and green dotted lines, respectively.}
        \label{var_alpha}
    \end{figure*}

Using the same AGN sample and control matching scheme, we recompute $\Delta\Sigma_\text{mol, inner}$, $\Delta f_\text{gas, inner}$, and $\Delta \nabla f_\text{gas}$ using a variable $\alpha_\text{CO}$ for the inner region. The results are presented in Figure \ref{var_alpha}. Using the $\alpha_\text{CO}$ prescription from \citet{Schinnerer24} produces similar distributions in these three key parameters and the median values of $\Delta\Sigma_\text{mol, inner}$, $\Delta f_\text{gas, inner}$, and $\Delta \nabla f_\text{gas}$ for the AGN population (reported as pink in the top right corner) are equal to the constant $\alpha_\text{CO}$ analysis within error.

    We also tested how the variable $\alpha_\text{CO}$ prescription would affect the different AGN classes (see Fig. \ref{delta_fgas_agn_split}) and found that the 1.5$\sigma$ and 1.8$\sigma$ signals of systematically lower $f_\text{gas, inner}$ in K01 and wAGN hosts, respectively, become more significant: 1.9$\sigma$ and 2.3$\sigma$, respectively. However, there is no added significance to any of the trends in Figure \ref{agn_trends}.